\documentclass[12pt]{article}       
\usepackage{authblk}
\usepackage{graphicx}
\usepackage{amssymb}
\usepackage{amsmath}
\usepackage{color}
\usepackage{mathtools}
\allowdisplaybreaks
\usepackage{multicol}
\usepackage{multirow}
\usepackage{tikz}
\usepackage{lscape}
\usepackage{pgfplots}
\usepackage{caption}
\usepackage[section]{placeins}
\usetikzlibrary{shapes,arrows}
\usepackage{array}
\usepackage{tikz}
\usetikzlibrary{shapes,arrows,backgrounds,patterns,calc}
\usetikzlibrary{decorations.pathreplacing}
\usepackage{xparse}
\tikzset{myarrow/.style={draw,fill=black,double arrow,minimum height = 3.5cm,} } 

\tikzstyle{every picture}+=[remember picture]
\newcommand{\tikzmark}[1]{\tikz[overlay,remember picture] \node (#1) {};}

\newcommand*{\BraceAmplitude}{0.4em}%
\newcommand*{\VerticalOffset}{0.5ex}%
\newcommand*{\HorizontalOffset}{1.0em}%
\NewDocumentCommand{\InsertLeftBrace}{%
    O{} 
    O{\HorizontalOffset,\VerticalOffset} 
    m   
    m   
    m   
}{%
    \begin{tikzpicture}[overlay,remember picture]
        \coordinate (Brace Top)    at ($(#4 |- #3.north) + (#2)$);
        \coordinate (Brace Bottom) at ($(#4 |- #5.south) + (#2)$);
    \draw [decoration={brace, amplitude=\BraceAmplitude}, decorate, thick, draw=blue, #1]
    (Brace Bottom) -- (Brace Top) ;
    \end{tikzpicture}%
}%
\tikzset{myarrow/.style={draw,fill=black,double arrow,minimum height = 3.5cm,} } 
\makeatletter
  {\def\@captype{figure}}
  {}
\newcommand{\diff}[2]{\frac{\partial#1}{\partial#2}}
\newcommand{\dif}[2]{\frac{d\,#1}{d\,#2}}
\newcommand{\be}{\begin{equation}}
\newcommand{\ee}{\end{equation}}
\newcommand{\bal}{\begin{align}}
\newcommand{\eal}{\end{align}}

\newcommand{\ol}{\overline}

\begin{document}

\title{A Stochastic Closure for Two-Moment Bulk  Microphysics of Warm Clouds:  Part II, Validation}

\author{David Collins\thanks{Corresponding author, email: \texttt{davidc@uvic.ca} } ~and Boualem Khouider\thanks{Tel. 250-721-7439, email: \texttt{khouider@uvic.ca} } }
\affil{               University of Victoria,  
               PO BOX 3060 STN CSC,  \\
                Victoria, BC,  
                 Canada V8W 3P4			}



\date{8 May 2016 \\ Submitted on 2 May 2016 to \\ \textbf{Meteorology and Atmospheric Physics}}

\maketitle

\begin{abstract}
The representation of clouds and associated processes of rain and snow formation remains one of the major uncertainties in climate and weather  prediction models. In a companion  paper (Part I), we systematically  derived a two moment bulk cloud microphysics model for collision and coalescence in warm rain based on the kinetic coalescence equation (KCE) and used stochastic approximations to close the higher order moment terms, and do so independently of the collision kernel. Conservation of mass and consistency of droplet number concentration of the evolving cloud properties were combined with numerical simulations to reduce the parametrization problem to three key parameters.   

Here, we constrain these three parameters based on the physics of collision and coalescence resulting in a ``region of validity.''  Furthermore, we theoretically validate the new bulk model by deriving a subset of the ``region of validity" that contains stochastic parameters that skillfully reproduces an existing model based on an a priori droplet size distribution by Seifert and Beheng (2001).  The stochastic bulk model is empirically validated against this model,  and parameter values that faithfully reproduce detailed KCE results are identified. 

Furthermore, sensitivity tests indicate that the stochastically derived model can be used with a time step as large as 30 seconds without significantly compromising accuracy, which makes it very attractive to use in medium to long range weather prediction models. These results can be explored in the future to select parameters in the ``region of validity" that are conditional on environmental conditions and the age of the cloud.

\end{abstract}

\section{Introduction}\label{sec:Intro}
Climate and numerical weather prediction models are large computer codes based on a discretization of the governing equations for fluid dynamics and thermodynamics for the coupled atmosphere-ocean-earth system.  Due to limitations in computing resources only a small part of the spectrum of scales involved in this interactive complex system are explicitly modelled and many small scale processes are represented via various recipes known as parameterizations.  The parameterization of cloud and precipitation processes constitute a major uncertainty in climate and weather prediction models.  For example, reliable determinations of the onset of rainfall and the radar reflectivity in clouds are crucial for the accuracy of models that rely on the parameterizations of cloud microphysical processes \cite{HP97}.  A large size gap exists between the length scales on which the cloud microphysical processes occur and the grid cell size of global and region climate models  \cite{BD12}.  Bulk parameterizations are simple equations that represent these processes in a computationally efficient manner \cite{CF08}. 

In warm clouds, bulk parameterizations can include terms representing nucleation, collision and coalescence, and precipitation \cite{WG13}.  The simplest models evolve an initial droplet size distribution (DSD) using only collision and coalescence terms: self-collection (cloud and rain), autoconversion, and accretion.  Auto conversion involves droplets of a size that are more strongly influenced by turbulence then the other collision and coalescence processes  \cite{RS03}, and it is the only process that is present in all of the equations that evolve both rain and cloud aggregates.  In a bulk parameterization of warm clouds, for example, a common technique was to derive or `fit' the auto conversion term for rain mixing ratio to data, and then make the remaining auto conversion terms to be functions of the rain mixing ratio term \cite{MK00,AS01,CF08}.  In Part I, we broke with that tradition by independently deriving all auto conversion terms while preserving conversation of mass and accuracy of number \cite{DC16a}.  The independent derivations of all auto conversion terms permitted further control of the auto conversion parameter.   This new technique is based on a systematic partitioning of the droplet distribution spectrum into cloud and rain aggregate means plus sets of point-wise random fluctuations. This stochastic  representation, which is in essence a new way of taking into account uncertainty in the DSD, is used as a basis to close high order terms to derive a two moment parameterization for bulk cloud microphysics directly from the kinetic coalescence equation (KCE).
 
Kessler is widely regarded as the first to parameterize a collision and coalescence process in a cloud microphysical context.  In a 1969 paper he used a Heaviside function to terminate autoconversion when a critical threshold of cloud liquid water content $L_c$ was reached \cite{YL04}.  Since then there have been many parameterizations; some continuing to only model auto conversion \cite{YL04,YL06} and others modelling a full suite of one and two moment bulk parameterizations \cite{MK00,AS01,CF08,WG13}.  The development of an auto conversion only model included one that softened the termination of auto conversion \cite{HS78}.  Others introduced a dependence on number concentration \cite{YL04} and \cite{YL06}.  A few bulk parameterizations have used fitted models, one to DNS \cite{CF08} and another to detailed schemes \cite{MK00}.  Similarly to the new stochastic model, Seifert and Beheng (2001, hereafter SB01) used the kinetic collection equation (KCE) to derive a two-moment bulk cloud parameterization of collision and coalescence in the form of an ODE system which evolves the mixing ratios and number concentrations of both cloud and rain aggregates  \cite{AS01}.   
The KCE is an integro-differential equation which evolves the droplet spectrum by summing over all possible collisions leading to two integrals modelling respectively the gain and loss of droplets of every possible size:  
\begin{equation}
\begin{aligned}  
  \diff{n(x,t)}{t} = \frac12\int_{0}^x n(x-x',t)n(x',t) K(x-x',x')dx' - \\  \int_0^{\infty} n(x,t)n(x',t)K(x,x')dx' \label{eq:KCE}
\end{aligned}
\end{equation} 
where $x$ is the drop mass, $n(x,t)$ is the number concentration, a density function, and $K(x,x')$ is the collision-coalescence kernel so that $n(x,t)n(x',t)$ $K(x,x')$ is the rate-density of collision-coalescence between two droplets of mass $x$ and $x'$, respectively.   

Khairoutdinov and Kogan (2000) \cite{MK00}, Seifert and Beheng (2001) \cite{AS01}, and Franklin (2008) \cite{CF08}, derived expressions for the auto-conversion process, and for the accretion process, as they affected rain mixing ratio. They used conservation of mass to deliver expressions for loss of cloud mixing ratio due to these processes, and used the equation  for mean droplet mass,
\begin{equation} \label{eq:xqN} 
\overline x = \frac{q}{N} ,
\end{equation} 
to deliver expressions for other terms in their parameterizations.  The auto-conversion parameterizations for the remaining evolved quantities ($N_c$, $N_r$, $q_c$) are simply functions of the rain mixing ratio terms.  

Deriving a model independent of the DSD is important because any assumed distribution may not be universally appropriate for differing cloud types \cite{IZ94,RW05b}.  Thus a climate model utilizing only the new stochastic bulk model as a microphysical parameterization scheme, could represent different cloud types, ages, and turbulent phenomena or intensities by simply selecting appropriate stochastic parameter values to be used in each climate model cell.

 As already mentioned, in lieu of applying any particular droplet size distribution (DSD), the stochastic bulk parameterization represented number concentration density and mixing ratio density as the sum of the state space mean and a point-wise fluctuation, and used a 2D Taylor expansion to represent the collision kernel \cite{DC16a}.  Further simplifications and a significant reduction in the number of parameters was achieved in the light of detailed numerical simulations based on the KCE by invoking two independent order of magnitude arguments: (i) the mean and standard deviation of the stochastic processes, and (ii) the temporal fluctuations of the evolving quantities  \cite{DC16a}.  Moreover, the resulting two-moment bulk cloud parameterization is independent of the underlying collision kernel and yet is consistent with existing models \cite{MK00,AS01,CF08}, and in particular the SB01 model can be recovered exactly with an appropriate choice of parameters.  The new stochastic model also independently derived auto-conversion and accretion terms for each of the four evolved quantities while preserving conservation of mass and consistency of number  \cite{DC16a}.  This new stochastic model is, to the authors knowledge, unique among existing stochastic models in that it is a stochastic differential equation-based method (`SDE-based' stochastic method) rather than being based on sampling an assumed probability distribution (`sample-based' stochastic method).  See Part I for a brief list of `sample-based' stochastic methods used to model cloud microphysical processes \cite{DC16a}.

The paper is organized as follows. In Section \ref{sec:MSBRP}, we recall the stochastic bulk parameterization model equations and use physical constraints associated with the processes of collision and coalescence to constrain the parameters and define a ``region of validity."  In Section \ref{sec:Param}, we apply the piece-wise polynomial kernel used in \cite{AS01} to the stochastic bulk parameterization and conduct validation tests based on detailed simulations of KCE. In particular, we do a term by term matching of the SB01 and the stochastically derived model equations and define the region in parameter space delimited by the applicability of the SB01 model. The ``region  of validity," in parameter space, based on collision and coalescence physics, and the region defined by the SB01 model are not identical but they are not disjoint either. A thorough exploration of these regions in parameter space is conducted for two typical initial cloud distributions, one representing a polluted/continental cloud and one representing a clean/oceanic cloud and the sets of parameters that reproduce the KCE results are identified for each distribution. As we will see in Section  \ref{sec:NumSim}, while the KCE results are reproduced  by a set of parameters within the intersection of  the two regions for the case of a clean cloud, in the case of a polluted cloud the matching parameters are found outside the SB01 model limits. A concluding discussion is given in Section \ref{sec:Conclu}.

\section{Controlling the Autoconversion Parameter} \label{sec:MSBRP}
In Part I, we systematically decomposed the number concentration density and mixing ratio density into state-space means and sets of point-wise fluctuations within cloud and rain droplet aggregates, summed the fluctuations over defined intervals and considered each instantaneous sum to be a point in a stochastic process.  Closure of the kinetic collection equation was achieved by representing higher moments as functions of the point-wise fluctuations.  The stochastic processes were defined as a mean and a standard deviation, and order of magnitude arguments eliminated the standard deviation terms.  The means of the stochastic processes are parameters.  Consistency of number and conservation of mass were used to reduce the degrees of freedom of the stochastic parameterization.   
Number concentration and mixing ratio are related by Equation \ref{eq:xqN}.
Writing the four stochastic bulk rate equations using three independent parameters yields
\begin{equation} 
\begin{aligned} \label{eq:threedegreesSBRP}
\dif{{N_c}}{t} = & -\frac14\left(3-2 \mu_{1n} \right)K_{cc}{N_c^2}-K_{cr}{N_cN_r}   \\
\dif{{N_r}}{t} = & ~  \frac{1}4 \left(1 - 2 \mu_{1n} \right) K_{cc}{N_c^2}   -  \left(\frac34- \mu_{4nc} \right) K_{rr}{N_r^2}   \\
\dif{{q_c}}{t} = &  -\left(\frac12-  \mu_{1m}  \right) K_{cc}{q_cN_c} - K_{cr}{q_cN_r} \\
\dif{q_r}{t} = & ~ \left(\frac12- \mu_{1m} \right) K_{cc} {q_cN_c}  + K_{cr}{q_cN_r}   
 \end{aligned}
\end{equation}
The physical meanings and effects of these parameters are
\begin{align}
\mu_{1n}   \hspace{1.0cm}  & \text{ controls the strength of cloud self-collection relative to auto-conversion }   \nonumber \\ 
\mu_{1m}   \hspace{1.0cm} &  \text{ controls the strength of auto-conversion }  \nonumber \\  
\mu_{4nc}  \hspace{1.0cm} & \text{ controls the strength of rain self-collection }  \nonumber \\   \nonumber
\end{align}  
The derivations and details which resulted in the stochastic parameterization presented in Equation \ref{eq:threedegreesSBRP} are given in the first paper.  The values of each of the three stochastic parameters could be acquired from the results of a bin-pair interaction detailed method.  

The `SDE-based' stochastic model is general enough for any kernel, but here we validate it by applying the piece-wise polynomial kernel given in Equation \ref{eq:SB_kernel}, and graphically acquire values of the stochastic parameters after constraining the stochastic parameters using conservation of mass, consistency of number, and the definition of auto conversion.  We compare it to sievert and Beheng's model using Bott's detailed method as a benchmark.  The values for results driven by a hydrodynamic kernel and by two turbulent kernels are presented in a third paper.


\subsection{Physical Limits: Consistency of Number and Conservation of Mass} \label{sec:connumber_conmass}
Since the kernel ($K_{cc}$) and number concentration ($N_c$) are non-negative, they are omitted from the following expressions.  Cloud self-collection and autoconversion each decrease cloud number concentration, thus: $\left(\frac34-\frac12 \mu_{1n} \right) \ge 0$.  Auto-conversion will cause rain number concentration to increase, thus:   $\frac{1}4\left(1- 2\mu_{1n} \right)  \ge0$.  Both of these conditions result in $\mu_{1n} \le 1/2$.  Because cloud self-collection does not affect rain number concentration, the cloud loss will be at least twice the rain increase: $ \left(\frac34-\frac12 \mu_{1n} \right) \ge  2(\frac{1}4\left(1- 2\mu_{1n} \right))$.  This results in $-1/2 \le \mu_{1n} $.  Auto conversion will decrease cloud mixing ratio and cause rain mixing ratio to increase, thus: $ \left(\frac12- \mu_{1m} \right)  \ge 0$.  Thus we have
\begin{equation} \label{eq:NC_cov_bounds}
-\frac12 \le \mu_{1n} \le \frac12 ~~~ \text{ and } ~~~ \mu_{1m} \le \frac12.
\end{equation}
The bounds on the two parameters related to cloud self-collection and auto conversion due to accuracy of number and conservation of mass can be represented on a single graph shown in the picture on the left in Figure \ref{fig:Combined_cov_graph}.  The lower bound on $\mu_{1m}$, or the strength of auto-conversion, is dependent on the distance of the mean cloud radius to the threshold radius:  the closer they are together the stronger the auto conversion rate and a smaller mean cloud radius demands a weaker auto-conversion rate.  This specific relationship can be quantified by examining the coefficients of the $K_{cc}$ terms in Equation \ref{eq:threedegreesSBRP}.

Additionally, rain self-collection decreases rain number concentration and is constrained by the following inequality: 
\begin{equation} \label{eq:SC_Param}
\mu_{4nc} \le\frac34 \hspace{0.5cm} \nonumber
\end{equation}

\subsection{Autoconversion and the Threshold Droplet Size} \label{sec:auto_thresholdsize}
There is a need to constrain the auto conversion parameter \cite{RW05b}.  Franklin (2007) and Khairoutdinov and Kogan (2000) have constrained it in an ad-hoc manner using best fit methods to DNS and to a detailed, bin based method, respectively.  Neither their constraints nor their parameters have physically based meanings \cite{CF07,MK00}.  In the stochastic bulk parameterization, the auto conversion parameter is further constrained using the relationship between the mean cloud radius and the threshold radius, and the definition of autoconversion; namely, that pre-collision, and post-collision, droplets must be smaller than and larger than the threshold radius, respectively.  All of the mass in an auto conversion collision moves from cloud to rain.  
\begin{center}
\begin{figure}
	\begin{tikzpicture}[thick,scale=1.25, every node/.style={scale=0.80}]
		\draw [<->, thick] (0,-2) -- (0,2);
		\draw [<->, thick] (-2,0) -- (2.0,0);
		
		\draw [dashed] (-2,1) -- (2,1);
		\draw [dashed] (-1,-2) -- (-1,1.5);
		\draw [dashed] (1,-2) -- (1,2);
		\path [fill=red!35] (-1.0,1.0) -- (1.0,1) -- (1,-1) -- (0.25,-1.25) -- (-0.25,-0.75) -- (-1,-1) -- (-1,1) ;
		
		\node [left] at (0,1.7) {$\mu_{1m}$};
		\node [below] at (1.7,0.6)  {$\mu_{1n}$};
		
		\node [right] at (1.80,1.05) {$\frac12$};
		\node [above] at (-1.35,-0.75) {$-\frac12$};
		\node [above] at (1.35,-0.75) {$\frac12$};
		\node [right] at (-0.55,1.15) {\footnotesize{no auto}};
		\node [right] at (0.95,-1.20) {\footnotesize{max}};
		\node [right] at (0.95,-1.45) {\footnotesize{SC}};
		\node [right] at (-1.65,-1.20) {\footnotesize{no}};
		\node [right] at (-1.65,-1.45) {\footnotesize{SC}};			%
		
\draw [<->, thick] (5,-3.2) -- (5,2);
\draw [<->, thick] (3,0) -- (7,0);

\draw [dashed] (3,1.0) -- (7,1.0);
\draw [dashed] (4,-3.15) -- (4,1.75);
\draw [dashed] (6,-3.15) -- (6,1.75);
\draw [dotted] (5,-3) -- (6,1);
\draw [dotted] (4,-3) -- (6,1);
\path [fill=red!25] (5.0,-3.0) -- (6,1) -- (4,-3) -- (4.33,-2.8) -- (4.66,-3.2) -- (5,-3) ;

\node [left] at (5.0,1.7) {$\mu_{1m}$};
\node [below] at (7.2,0.6)  {$\mu_{1n}$};

\node [right] at (6.8,1.05) {$\frac12$};
\node [below] at (6.25,-0.0) {$\frac12$};
\node [below] at (3.7,-0.0) {$-\frac12$};
\node [below] at (3.7,-2.75) {$-\frac32$};
\node [below] at (3.50,-0.70) {$-\frac12$};
\node [right] at (4.45,1.15) {\footnotesize{no auto}};
\node [right] at (5.95,-1.20) {\footnotesize{max}};
\node [right] at (5.95,-1.45) {\footnotesize{SC}};
\node [right] at (3.35,-1.70) {\footnotesize{no}};
\node [right] at (3.35,-1.95) {\footnotesize{SC}};

\node [above] at (6.75,-2.75) {$\frac{x^*}{\overline{x}_c}=4$};
	\end{tikzpicture} 
	\caption[Constraining the Auto Conversion Parameter in the $\mu_{1n}$-$\mu_{1m} $ plane]{Physical constraints (conservation of mass and accuracy of number) on the stochastic parameters associated with cloud self-collection and with auto-conversion.  The triangle in the figure on the right extends to $(-\frac12,-\frac72)$, and is representative of the situation when the threshold radius is 40 $\mu$m and the mean cloud radius is 25.2 $\mu$m.   The mean cloud radius $\ol r_c$ and mean cloud mass $\ol x_c$ are related by $\ol x_c = 4/3 \pi \rho \ol r_c^3$, and similarly for the threshold mass and radius.\label{fig:Combined_cov_graph} }
\end{figure}
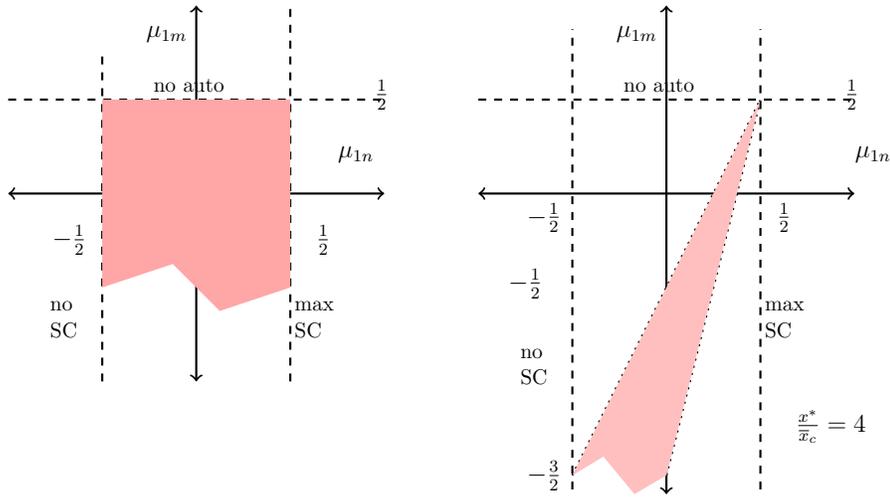
\end{center}
The effect that auto-conversion has on the rain spectrum is described by the first terms on the rhs side of the second and fourth equations in Equation \ref{eq:threedegreesSBRP}.  These are related to number and mixing ratio, respectively.  Therefore, given $\frac{q_c}{N_c}=\ol{x}_c$, the equation requiring that post-collision droplets have a mass greater than the mass of the threshold droplet is
\be
x^* \le \frac{(\frac12-\mu_{1m})K_{cc}q_cN_c}{\frac14(1-2\mu_{1n})K_{cc}N_c^2} ~ = ~  \frac{(\frac12-\mu_{1m})}{\frac14(1-2\mu_{1n})} \ol{x}_c 		\nonumber
\ee
Letting $\beta = \frac{x^*}{\ol x_c}$ and solving for $\mu_{1m}$ gives the equation for the upper boundary of the triangular region in the graph on the right in Figure \ref{fig:Combined_cov_graph}:
\be
\mu_{1m} \le \frac{2-\beta}4 + \frac{\beta}2\mu_{1n}
\ee
The lower boundary is constructed from the requirement that pre-collision droplets are each smaller than the threshold droplet.  This is accomplished algebraically by considering that the number of cloud droplets lost to auto-conversion is twice the number rain droplets produced by the same process.  The number of cloud droplets lost is given by doubling the first term on the rhs if the second differential equation in Equation \ref{eq:threedegreesSBRP} and proceeding similarly as for the upper boundary.
\be
x^* \ge \frac{(\frac12-\mu_{1m})K_{cc}q_cN_c}{\frac12(1-2\mu_{1n})K_{cc}N_c^2} ~ = ~  \frac{(\frac12-\mu_{1m})}{\frac12(1-2\mu_{1n})} \ol{x}_c 		\nonumber
\ee
which yields the lower bound on the triangular region in the graph on the right in Figure \ref{fig:Combined_cov_graph}:
\be
\mu_{1m} \ge \frac{1-\beta}2 + \beta \mu_{1n}
\ee

Using $\beta=4$ as is depicted in the right hand graph in Figure \ref{fig:Combined_cov_graph}, we get for the lower and upper bounds, respectively:
\be
\mu_{1m} \ge -\frac32 + 4 \mu_{1n} ~~~ \text{ and } ~~~ \mu_{1m} \le -\frac12 + 2\mu_{1n} 		\nonumber
\ee
When a cloud is young or heavily polluted, we can expect $\beta$ to be large and the values for the stochastic parameters to be near the upper right corner of the region in the $\mu_{1n}-\mu_{1m}$ plane.  This demands strong cloud self-collection and weak auto-conversion, which is to be expected when the cloud droplets are primarily small.  By contrast in a mature cloud or one with primarily large droplets, $\beta$ will be smaller and the region giving values for $\mu_{1n} ~\& ~ \mu_{1m}$ will be different.  

The strength of cloud self-collection is determined from the relationship between the first terms in the first two differential equations in Equation \ref{eq:threedegreesSBRP}.  When the value of $\mu_{1n}$ is chosen such that the first term in the equation for $N_r$ is zero (i.e. $\mu_{1n}=\frac12$), then none of the  droplets subtracted in the equation for $N_c$ creates rain droplets and thus all of them create cloud droplets and cloud self collection is maximized.  Consequently, auto-conversion in non-existent and the region of valid values for $\mu_{1n} ~\& ~ \mu_{1m}$ vanishes at the intersection of the lines for max cloud self-collection and no auto-conversion (i.e. $\mu_{1n}=\frac12$ and $\mu_{1m}=\frac12$).  

Alternately, when the value of $\mu_{1n}$ is chosen such that all of the droplets subtracted in the evolution of $N_c$ in Equation \ref{eq:threedegreesSBRP} produce rain droplets in the second equation, $N_r$ (i.e. $\mu_{1n}=-\frac12$), then cloud self-collection is non-existent and auto conversion is maximized.  In other words, the coefficient of the first term in $N_c$ is twice the coefficient of the first term in $N_r$ because two cloud droplets produce one rain droplet in the process of auto-conversion.  

\subsection{Constraint of the Autoconversion Parameter}
The constraining of the auto conversion parameter is possible because the four auto conversion terms are derived independently.  The independent derivations allow the ratios of mixing ratio and number to be compared to the threshold mass.  This is not possible when one auto conversion term is a function of another auto conversion term.  The equations for the two sloped sides of the triangle in Figure \ref{fig:Combined_cov_graph} appear because of the independent derivations of the number concentration and mixing ratio equations.  The third side is resultant from the two cloud droplet terms for number being derived independently.  

Conservation of mass and consistency of number create the same bounds as the upper right vertex of the triangle in Figure \ref{fig:Combined_cov_graph} which is the location on the graph representative of strong cloud self-collection and weak (to non-existent) auto conversion.  Stochastic parameter values in this area of the triangle model a polluted cloud with many small droplets.  Consistency of number constrains the maximal and minimum value of the cloud self-collection parameter.  Conservation of mass only constrains the maximal value (weakest strength) of the auto conversion parameter.  

The maximum strength of auto conversion is limited by the lower vertex of the triangle.  Auto conversion is maximal when the mean cloud radius is close to the threshold radius which gives minimal values to $\mu_{1m}$ in the two vertices on the left side of the triangle which collaborates some of Franklin's results \cite{CF08}.  Alternatively, when cloud self-collection is strong, the lower bounds on the value of the auto conversion parameter is restricted further than the apparent bounds given by the perimeter of the triangle: a strong cloud self-collection parameter ($\mu_{1n} \approx 1/2-\epsilon$) disallows $\mu_{1m}$ taking values from the lower area of the triangle.  Thus $\mu_{1m}$ is constrained to be `near' to 0.5, i.e. modelling a weaker auto conversion rate.

\section{The Stochastic Model with a Piece-Wise Kernel} \label{sec:Param}
The stochastically derived model (\ref{eq:threedegreesSBRP}) is general and can be adapted for any kernel, but here we validate it for the case of the piece-wise polynomial kernel in (\ref{eq:SB_kernel}).  This particular kernel was used by SB01 to derive a similar set of two-moment equations based on the averaging the KCE with a specified droplet distribution to close high order terms.  For the sake of completeness, we also derive a region of validity for the stochastic model in (\ref{eq:threedegreesSBRP}) by matching it to the SB01 model based on their assumptions of parameters, namely, the shape parameter of the assumed distribution, $\nu$, and the time scale $\tau$ for auto-conversion and accretion processes.  We call this region the SB01 region of validity in contrast to the stochastic region of validity derived above based on the physical constraints motivated by the properties of the collision and coalescence process of cloud droplets.  We then use detailed simulations of KCE as benchmarks to find parameter values that are physical based on two measures (the time at which 50\% of cloud mass is converted into rain and the size of the mean rain droplet at that time) within the regions of validity of the stochastically derived model, with this specific piece-wise polynomial kernel, vis-a-vis the performance of the SB01 model.    The cases involving a hydrodynamic kernel and two turbulent kernels will be considered in a future publication. 




\subsection{Seifert and Beheng's Model (2001)}	\label{sec:SB01}

SB01 developed a parameterization that models the effects of collision and coalescence on the evolution of four bulk cloud microphysical properties.  Their parameterization was subsequently used in 2010 in a 1-D rain shaft model that included droplet breakup and condensation and evaporation \cite{AS10b}.  

 In \cite{AS01}   the centres of cloud and rain aggregates: $N_c\ol{x_c}=q_c$ and $N_r\ol{x_r}=q_r$ are systematically used, and a gamma distribution and $\ol{x}_c \ll x^* \approx \ol{x}_r$ are assumed as initial conditions to close the two-moment problem.  They derived auto-conversion and accretion expressions as functions of their tuning parameters: the gamma distribution shape parameter $0 \le \nu \le 3$ and a non-dimensional time parameter $0 < \tau \le 1$:
\begin{equation} 
\begin{aligned}  \label{eq:SB01autoaccr} 
\Gamma_{\text{auto}} ~ = ~ & \frac{1}{20}\frac{(\ol{x_c})^2}{x^*}\frac{(\nu+2)(\nu+4)}{(\nu+1)^2} \left[1+\frac{\Phi_{\text{au}}(\tau)}{(1-\tau)^2}\right]  \\
\Gamma_{\text{accr}} ~ = ~ & \Phi_{\text{ac}}(\tau)\left(1+\frac{\nu+2}{\nu+1}\frac{\ol x_c}{\ol {x_r}} \right)   \\
 \end{aligned}
\end{equation}
where $x^*$ is a threshold mass corresponding to a radius of $40\mu$m.  The functions $\Phi_{\text{ac}}(\tau)=\left( \frac{\tau}{\tau+0.0005}\right)^4$ and $\Phi_{\text{au}}(\tau)=600 \tau^{0.68}\left(1-\tau^{0.68}\right)^3$ represent internal time scales for the corresponding processes that are derived empirically using simulated data with $\tau=1-q_c/(q_c+q_r)$ \cite{AS01}. 

SB01\cite{AS01} used a piece-wise combination of Golovin's sum of mass kernel and Long's quadratic polynomial kernel, referred to here simply as the piece-wise polynomial kernel.
\be \label{eq:SB_kernel}
K(x,y)=
\begin{cases}
k_c(x^2+y^2), \hspace{0.5cm} x \text{ and } y < x^* \\
k_r(x+y), \hspace{0.8cm} x \text{ or } y \hspace{0.25cm} \ge x^*
\end{cases} 
\ee
where $k_c = 9.44 \times 10^9$ cm $^3$g$^{-2}$s$^{-1}$, and $k_r = 5.78 \times 10^3$ cm $^3$g$^{-1}$s$^{-1}$.  Using abbreviated notation, their bulk model is
\begin{equation} 
\begin{aligned}  \label{eq:SB01undermark}
\diff{N_c}{t} = & ~  - \underbrace{\frac{\nu+2}{\nu+1}k_cq_c^2 }_{\text{auto \& self-coll} } -  ~\underbrace{ \Gamma_{\text{accr}}k_rq_rN_c  }_{\text{accretion} }   \\
\diff{N_r}{t} = & ~ \underbrace{  \frac{\Gamma_{\text{auto}}}{x^*} k_cq_c^2 }_{\text{autoconversion} } ~ - ~ \underbrace{k_rq_rN_r  }_{\text{self-coll}}   \\
\diff{q_c}{t} = & ~ - ~ \underbrace{\Gamma_{\text{auto}}k_cq_c^2  }_{\text{autoconversion} } ~  - ~ \underbrace{\Gamma_{\text{accr}} k_rq_cq_r   }_{\text{accretion} }    \\
 \diff{q_r}{t} = & ~  \underbrace{\Gamma_{\text{auto}}k_cq_c^2  }_{\text{autoconversion} } ~ + ~ \underbrace{  \Gamma_{\text{accr}} k_rq_cq_r   }_{\text{accretion} }
 \end{aligned}
\end{equation}
The parameters $\nu$ and $\tau$ are used in seven of eight terms in their bulk parameterization.  

The terms in Equation \ref{eq:SB01undermark} are labeled by the collision processes that they model.  Auto conversion occurs in each of the four evolving quantities and accretion occurs in three equations.  The coefficients of seven of the eight terms are dependent on an ad-hoc shape parameter $\nu$, and all are dependent on one or more of six other parameters ($k_c$, $k_r$, $\tau$, $x^*$, $\ol x_c$, $\ol x_r$).  

Although SB01 omitted $\frac{\nu + 2}{\nu + 1}\frac{\ol x_c}{\ol x_r}$ from the accretion terms in their final parameterizations, this factor is included in $\Gamma_{\text{acct}}$ in Equation \ref{eq:SB01autoaccr} for completeness, and it appears in precisely the same terms in the stochastic model once the same piece-wise polynomial kernel is applied.  These eight coefficients, excluding the collision kernels, are equated with the coefficients of the corresponding terms in the stochastic parameterization.

\subsection{Reproduce Existing Parameterization} \label{sec:ReproduceExistParam}
We reproduce a bulk model driven by a piece-wise kernel.  Applying a different kernel, say a hydrodynamic or a turbulent one, would yield a different parameterization.  The piece-wise polynomial kernel shown in Equation \ref{eq:SB_kernel} is applied to the stochastic bulk parameterization Equation \ref{eq:threedegreesSBRP}.  This analytic kernel introduces the centroids of the cloud and rain aggregates into the stochastic parameterization which appear in the accretion terms.  The stochastic bulk parameterization of cloud microphysical collision processes driven by the Long-Golovin kernel is:
\begin{equation} 
\begin{aligned}  
\dif{{N_c}}{t} & = - \frac1{2} \left(3-2\mu_{1n} \right)k_{c}{q_c^2} - \left(1+\frac{\ol{x_c}}{\ol{x_r}}\right)k_{r}{q_rN_c}  \\
\dif{{N_r}}{t} & = ~ \frac1{2} \left(1- 2\mu_{1n} \right) k_{c}{q_c^2}   - \left(\frac32 - 2\mu_{4nc} \right) k_{r}{q_rN_r}   \\
\dif{{q_c}}{t} & = - \left(1-2\mu_{1m} \right)\ol x_c k_{c}{q_c^2}-\left(1+\frac{\ol{x_c}}{\ol{x_r}}\right)k_{r}{q_cq_r} \\
\dif{{q_r}}{t} & = ~ \left(1- 2\mu_{1m} \right) \ol x_c k_{c}{q_c^2}   + \left(1+\frac{\ol{x_c}}{\ol{x_r}}\right)k_{r} {q_cq_r}. \label{eq:MSSB}
\end{aligned}
\end{equation}
A term-by-term comparison of Equations ($\ref{eq:SB01undermark}$ and $\ref{eq:MSSB}$) reveals an exact match regarding the dependencies on aggregate number concentrations ($N_c$ and $N_r$) and mixing ratios ($q_c$ and $q_r$).  The coefficients of corresponding terms are equated to determine the range of the stochastic parameters as functions of the range of values that Seifert and Beheng gave to their parameters, $\tau$ and $\nu$.

\subsection{Select Stochastic Parameter Values}
We make a systematic comparison by matching corresponding terms.  In both Seifert and Beheng's parameterization and the stochastic parameterization, the auto conversion terms for cloud and rain mixing ratio are equal with the opposite sign.  The same is true for the cloud and rain accretion terms.   Consequently, with regard to mixing ratio, only the auto conversion and accretion terms that evolve rain mixing ratio are compared because the cloud terms have been restricted by conservation of mass.  The stochastic term for rain self-collection is compared with the corresponding term in the Seifert and Beheng parameterization.  The following calculations use a threshold radius of $40\mu$m and a mean cloud radius of $10\mu$m.  The auto conversion term for rain number concentration contains a stochastic parameter ($\mu_{1n}$) that is a different parameter than the one for mixing ratio ($\mu_{1m}$), and so the rain number coefficients are compared.  The first terms in both cloud number concentration equations combine the processes of auto conversion and cloud self-collection.  These are compared; however, they contain the same stochastic parameter ($\mu_{1n}$) as in rain number concentration.  The more meaningful range of values for $\mu_{1n}$ is derived from cloud number concentration.  The physically permissible values of the stochastic parameters that reproduce Seifert and Beheng's bulk parameterization are collected tabularly and displayed graphically in Section \ref{sec:SelectedValues}.  

\textbf{Auto conversion, mixing ratio} \label{sec:auto_compare} \\
The coefficients for auto conversion of rain (and cloud) mixing ratio are set equal  
\begin{equation}
\left(1- 2\mu_{1m} \right) \ol x_c ~=~  \frac{1}{20}\frac{(\ol{x_c})^2}{x^*}\frac{(\nu+2)(\nu+4)}{(\nu+1)^2} \left[1+\frac{\Phi_{\text{au}}(\tau)}{(1-\tau)^2}\right]  \nonumber
\end{equation}
and solved for the stochastic parameter:
\begin{equation}
\mu_{1m} ~=~ \frac12 - \frac{1}{40}\frac{\ol{x_c}}{x^*}\frac{(\nu+2)(\nu+4)}{(\nu+1)^2} \left[1+\frac{\Phi_{\text{au}}(\tau)}{(1-\tau)^2}\right]  \nonumber
\end{equation}
The shape parameter $\nu$ fraction is monotonic over the interval $(0,3)$, and the time parameter $\tau$ function has a maximum at $\tau\approx0.27$.  The bounds that Seifert and Beheng set for $\nu$ and $\tau$ and the maximum of $\tau$ are used to determine the range of the stochastic parameter governing the strength of auto conversion: 
\begin{equation}  \label{eq:rmLG}
0.20094 \le \mu_{1m} \le 0.49829.
\end{equation}
This range is displayed graphically (blue patterned region) in Figure \ref{fig:LG_values10um}.

\textbf{Auto conversion, rain number concentration}  \label{sec:AutoRNC}  \\
Equating the corresponding auto-conversion terms for rain number concentration gives
\begin{equation} 
\frac1{2} \left(1- 2\mu_{1n} \right) =  \frac{1}{20}\left(\frac{\ol{x_c}}{x^*}\right)^2\frac{(\nu+2)(\nu+4)}{(\nu+1)^2} \left[1+\frac{\Phi_{\text{au}}(\tau)}{(1-\tau)^2}\right].  \nonumber \\
\end{equation}
Solving for the stochastic parameter gives:
\begin{equation}    \label{eq:clLG}
\mu_{1n} ~=~  \frac12 - \frac{1}{20}\left(\frac{\ol{x_c}}{x^*}\right)^2\frac{(\nu+2)(\nu+4)}{(\nu+1)^2} \left[1+\frac{\Phi_{\text{au}}(\tau)}{(1-\tau)^2}\right] 
\end{equation}
The stochastic parameter governing the strength of cloud self-collection relative to auto conversion is constrained by Equation \ref{eq:clLG} to values less than but close to the maximum permissible value: 
$0.49066 \le \mu_{1n} \le 0.49997.$  The lower bound is consistent with the value of $\mu_{1n}$ when  $\mu_{1m}=0.20094$ given the constraint that a rain droplet produced during an auto conversion process must be larger than the threshold mass $x^*$.

\textbf{Auto conversion, cloud number concentration}  \label{sec:AutoCNC}  \\
Equating the corresponding auto-conversion terms for cloud number concentration gives:
\begin{equation} 
\frac12 (3-2\mu_{1n}) = \frac{\nu +2}{\nu +1}  ~~~\text{ which implies }  ~~~ \mu_{1n} = \frac32 - \frac{\nu + 2}{\nu + 1}  \nonumber
\end{equation}
For any value of $\nu$ specified by Seifert and Beheng ($0 \le \nu \le 3$), the bounds on the stochastic parameter is readily calculated: $0.25 \le \mu_{1n} \le 0.5$.  The derivation of their auto conversion term for rain umber concentration uses the assumptions that $\ol x_r \approx x^*$ and $\ol x_c \ll \ol x^*$.  These assumptions may be true during the early stages of a cloud's lifetime, but are less likely to be true in a mature cloud.  Therefore, the lower bound ($\mu_{1n}=0.25)$ derived in this section will be used to constrain this stochastic parameter.  Since the upper bound based on cloud number concentration terminates auto conversion, the upper bound ($\mu_{1n}=0.49997$) based on rain number concentration will be used.
\begin{equation}  \label{eq:cnLG}
0.25 \le \mu_{1n} \le 0.49997
\end{equation}
This range is displayed graphically (blue patterned region) in Figure \ref{fig:LG_values10um}.

\textbf{Accretion}  \\
The none of the three accretion terms in the stochastic bulk parameterization contain any stochastic parameters.  Except for the sign, the coefficients for the three accretion terms are identical:
\begin{equation}
 \left(1+\frac{\ol{x_c}}{\ol{x_r}}\right) ~=~  \Phi_{\text{ac}}(\tau)\left(1+\frac{\nu+2}{\nu+1}\frac{\ol x_c}{\ol {x_r}} \right) \nonumber
\end{equation}
The non-dimensional time function is an ad-hoc parameter that was included after completing their derivations, and the shape parameter is a by-product of the assumed droplet gamma distribution.  Seifert and Beheng removed the shape parameter function from the accretion terms in their final parmeterization scheme.  There are no arbitrary or ad-hoc parameters in the stochastic bulk rate parameterization.  Consequently, the only difference between the stochastic parameterization and their `adjusted' parameterization is the non-dimensional time function which was added after derivations were complete.  This function serves to slow accretion during the early and late stages of a cloud's lifetime and accelerate accretion during the middle stages, maximally when $\frac{q_r}{q_c+q_r} \approx 0.27$.  

A first order truncation of the collision kernel in the stochastic parameterization may give stochastic parameters in the accretion terms which may serve the same purpose as their non-dimensional time function.  However, all stochastic parameters are associated with concrete physical meanings.  Trading the higher order Taylor expansion terms of the collision kernel for the simplicity of the `SDE-based' stochastically derived set of coupled ODE's and omitting the ad-hoc non-dimensional parameter inclusions, all accretion terms are identical.

\textbf{Rain self-collection, number concentration} \label{sec:RainSC_NC}  \\
Equating the corresponding rain self-collection terms for number concentration gives:
\bal   \label{eq:rnLG}
\frac32- 2\mu_{4nc} = 1, ~~~ \text{ thus } ~~~ \mu_{4nc} = \frac14 < \frac34 ~~~ \text{as required.} 
\end{align}

\begin{center}
\begin{figure}[t]
\begin{center}
	\begin{tikzpicture}[thick,scale=1.25, every node/.style={scale=0.80}] 
		\draw [<->, thick] (0,-3.2) -- (0,2);
		\draw [<->, thick] (-2,0) -- (2.0,0);
		
		\draw [dashed] (-2,1) -- (1.85,1);
		\draw [dashed] (-1,-3.15) -- (-1,1.5);
		\draw [dashed] (1,-3.15) -- (1,1.5);
		\draw [dashed] (0.5,-0.5) -- (0.5,1.0);
		\draw [dotted] (-2,0.4) -- (1.85,0.4);
		\path [fill=red!25] (0.8750,-3.0) -- (1.0,1) -- (0.9375,-3) -- (0.9167,-3.01) -- (0.8959,-2.99) -- (0.8750,-3) ;
		\fill[pattern=north west lines, pattern color=blue, opacity=0.7]  (0.5,0.4) -- (0.5,1.0) -- (1.0,1.0) -- (1.0,0.4) -- (0.5,0.4)  ; 		
		\node [left] at (0,1.7) {$\mu_{1m}$};
		\node [below] at (1.8,0.0)  {$\mu_{1n}$};
		
		\node [right] at (1.80,1.00) {\footnotesize{0.5}};
		\node [below] at (0.5,-0.5) {\footnotesize{0.25}};
		\node [above] at (-1.0,1.5) {\footnotesize{-0.5}};
		\node [above] at (1.0,1.5) {\footnotesize{0.5}};
		\node [left] at (-1.0,-3.0) {-\footnotesize{1.5}};
		\node [left] at (-1.0,-1.0) {-\footnotesize{0.5}};
		\node [right] at (1.80,0.40) {\footnotesize{0.201}};
		\node [right] at (-0.55,1.15) {\footnotesize{no auto}};
		\node [right] at (1.00,-1.20) {\footnotesize{max}};
		\node [right] at (1.00,-1.45) {\footnotesize{SC}};
		\node [right] at (-1.45,-1.70) {\footnotesize{no}};
		\node [right] at (-1.45,-1.95) {\footnotesize{SC}};			
		
		\draw [dashed] (3,1.5) -- (6.8,1.5);
		\draw [dashed] (3.5,-2.75) -- (3.5,1.75);
		\draw [dashed] (6.5,-2.75) -- (6.5,1.75);
		\draw [dotted] (3,-2.5) -- (6.8,-2.5);
		\fill[pattern=north west lines, pattern color=blue, opacity=0.7]  (3.5,-2.5) -- (3.3,-1.2) -- (3.7,0.2) -- (3.5,1.5) -- (6.5,1.5) -- (6.5,-2.5) -- (3.5,-2.5)  ; 
		\path [fill=red!45, fill opacity=0.6] (4.312,-3.0) -- (6.5,1.5) -- (5.406,-3) -- (5.041,-3.1) -- (4.677,-2.9) -- (4.312,-3.0) ; 
		\node [left] at (3.5,-0.2) {\footnotesize{0.25}$\le \mu_{1n}$};
		\draw [<-, ultra thick, blue] (2.5,-0.4) -- (4.25,-0.4);
			
		
		\node [right] at (4.65,1.65) {\footnotesize{no auto}};
		\node [right] at (6.50,-0.45) {\footnotesize{max}};
		\node [right] at (6.50,-0.70) {\footnotesize{SC}};
		
		\node [below] at (3.5,-2.80) {\footnotesize{0.485}};
		\node [below] at (6.5,-2.80) {\footnotesize{0.50}};
		\node [left] at (7.3,1.5) {\footnotesize{0.5}};
		\node [right] at (6.75,-2.5) {\footnotesize{0.201}};
		
		\node [above] at (2.25,-2.75) {$\frac{x^*}{\overline{x}_c}=64$};
	\end{tikzpicture} 
	\caption[Region of Validity in $\mu_{1n}$-$\mu_{1m} $: Polluted Cloud]{The graph on the left is similar to and has the same scale as the two graphs in Figure \ref{fig:Combined_cov_graph}.  The lower bound of the triangle in the figures extends to $(-\frac12,-63.5)$, and is representative of the situation when the threshold radius is 40$\mu$m and the mean cloud radius is 10$\mu$m.   The graph on the right is a blow-up of the intersection of (i) the region (solid, red) of physically permissible values of stochastic parameters and (ii) the region (patterned, blue) of values that reproduce Seifert and Beheng's parameterization for their range of parameters.  \label{fig:LG_values10um} }
\end{center}	
\end{figure}
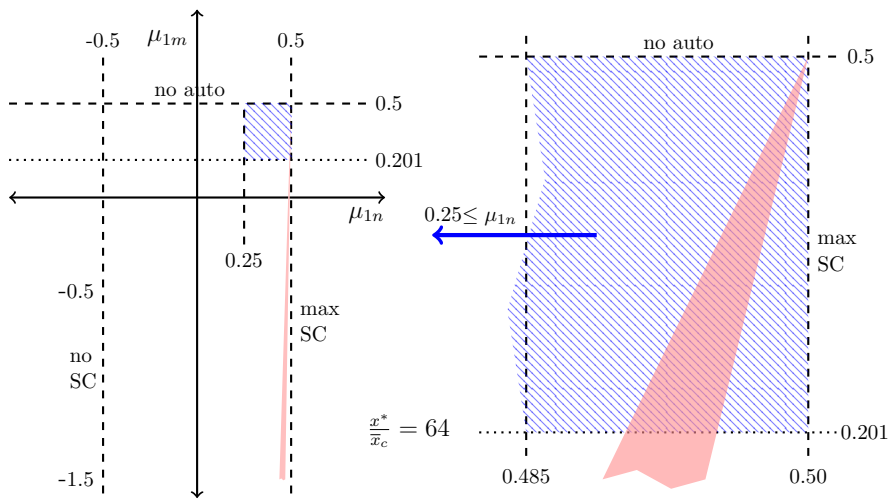
\end{center}

\subsection{Regions of Validity in Parameter Space}  \label{sec:SelectedValues}
The stochastic parameterization reproduces Seifert and Beheng's parameterization when $\mu_{1n}$, $\mu_{1m}$, and $\mu_{4np}$ are restricted to values identified in Equations ($\ref{eq:rmLG}$, $\ref{eq:cnLG}$, and $\ref{eq:rnLG}$).  These restrictions are formed by making the stochastic parameters to be functions of Seifert and Beheng's parameters: $\nu$ and $\tau$.  Equations ($\ref{eq:rmLG}$ and $\ref{eq:cnLG}$) bound the values of $\mu_{1n}$ and $\mu_{1m}$  and create a 2D space of parameter values which is displayed as the patterned (blue) in Figure \ref{fig:LG_values10um}.    Values within this region control the strength of cloud self-collection and auto conversion.  This rectangular region intersects the triangular (solid, red) region formed using the physical constraints from Sections \ref{sec:connumber_conmass} and \ref{sec:auto_thresholdsize}.

This intersection is a pentagonal region with vertices (counter-clockwise from lower left in the graph on the right of Figure \ref{fig:LG_values10um} at (0.49066, 0.20094), (0.49533, 0.20094), (0.49997, 0.49808), (0.49997, 0.49829), and (0.49995, 0.49829).  This pentagon is most clearly depicted in Figure \ref{fig:Contour_10um} when discussing sensitivity in Section \ref{sec:SensitivityTests}.  The bounds of values of $\mu_{1n}$ and $\mu_{1m}$ in this pentagonal region are listed in Table \ref{tab:ST_param10um}.  Three of the sides of the pentagon are apparent in the graph on the right in Figure \ref{fig:LG_values10um}.  The graphs in Figure \ref{fig:LG_values10um} and the values in Table  \ref{tab:ST_param10um} are derived using a threshold radius of $40\mu$m and an initial mean cloud radius of $10\mu$m which is representative of a terrestrial or `polluted' cloud.   The equations for the two sloped sides of the solid, red triangle in Figure \ref{fig:LG_values10um} are given as upper and lower bounds of $\mu_{1m}$ in Table \ref{tab:ST_param10um}.

\begin{table}[h!]
\begin{center}
\begin{tabular}{ c >{\centering\arraybackslash}m{1.75cm}  >{\centering\arraybackslash}m{5.5cm} >{\centering\arraybackslash}m{1.75cm} }
\hline
Parameter & Lower  & Condition & Upper  \\ 
\hline \hline
$\mu_{1n}$ & 0.49066 & --- & 0.49997 \\

\multirow{3}{*}{$ \mu_{1m} $ \tikzmark{Center Mark}} & \tikzmark{Top Mark}0.20094 & $0.49066 \le \mu_{1n} \le 0.49533$  &  \multirow{2}{*}{32$\mu_{1n}$-15.5} \\

 & \multirow{2}{*}{64$\mu_{1n}$-31.5}  & $0.49533 \le \mu_{1n} \le 0.49995$  &  \\
 
  & \tikzmark{Bottom Mark} & $0.49995 \le \mu_{1n} \le 0.49997$  & 0.49829 \\
 
 $\mu_{4np}$ & \multicolumn{3}{ c }{=0.25}    \\

\hline
\end{tabular} 
\InsertLeftBrace[black, ultra thick]{Top Mark}{Center Mark}{Bottom Mark}
\caption[Stochastic Parameter Bounds: Polluted Cloud]{The bounds of the auto conversion $\mu_{1m}$ and cloud self-collection $\mu_{1n}$ parameters for simulations with a threshold radius of 40$\mu$m and a mean cloud mass of 10$\mu$m. \label{tab:ST_param10um} }
\end{center}
\end{table}

For the `clean' cloud, an initial mean cloud radius of $16\mu$m is used with the same threshold radius.  The bounds of the stochastic parameters for a `clean' or marine cloud when the evolution is driven by the Long-Golovin kernel is shown graphically in Figure \ref{fig:LG_values16um} and numerically in Table \ref{tab:ST_param16um}.   In a `clean' cloud, the ratio $\frac{\ol{x}_c}{x^*}$ forces a sufficiently smaller upper bound on $\mu_{1n}$ (weaker cloud self-collection and stronger auto conversion) such that the intersecting region is a quadrilateral.  The vertices (counter-clockwise from lower left in the graph on the right of Figure \ref{fig:LG_values16um} of this four-sided figure are (0.34320, -0.72496), (0.4216,  -0.72496), (0.49300, 0.39063), and (0.49300, 0.44531).   Three of the sides of the quadrilateral are apparent in Figure \ref{fig:LG_values16um}.  This quadrilateral is most clearly depicted in Figure \ref{fig:Contour_16um} when discussing sensitivity in Section \ref{sec:SensitivityTests}.  The equations for the two sloped sides of the solid, red triangle are given as upper and lower bounds of $\mu_{1m}$ in Table \ref{tab:ST_param16um}.
\begin{center}
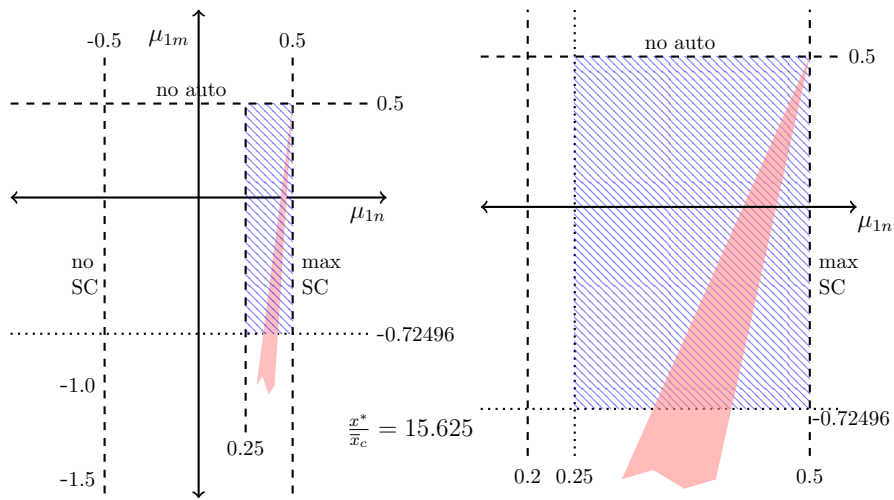
\begin{figure}[t]
\begin{center}
	\begin{tikzpicture}[thick,scale=1.25, every node/.style={scale=0.80}] 
		\draw [<->, thick] (0,-3.2) -- (0,2);
		\draw [<->, thick] (-2,0) -- (2.0,0);
		
		\draw [dashed] (-2,1) -- (1.85,1);
		\draw [dashed] (-1,-3.15) -- (-1,1.5);
		\draw [dashed] (1,-3.15) -- (1,1.5);
		\draw [dashed] (0.5,-2.5) -- (0.5,1.0);
		\draw [dotted] (-2,-1.45) -- (1.85,-1.45);
		\path [fill=red!25] (0.8750,-3.0) -- (1.0,1) -- (0.8080,-2) -- (0.7440,-2.10) -- (0.6800,-1.90) -- (0.6160,-2)  -- (1.0,1)  ;
		\fill[pattern=north west lines, pattern color=blue, opacity=0.7]  (0.5,-1.45) -- (0.5,1.0) -- (1.0,1.0) -- (1.0,-1.45) -- (0.5,-1.45)  ; 		
		\node [left] at (0,1.7) {$\mu_{1m}$};
		\node [below] at (1.8,0.0)  {$\mu_{1n}$};
		
		\node [right] at (1.80,1.00) {\footnotesize{0.5}};
		\node [below] at (0.5,-2.5) {\footnotesize{0.25}};
		\node [above] at (-1.0,1.5) {\footnotesize{-0.5}};
		\node [above] at (1.0,1.5) {\footnotesize{0.5}};0.5
		\node [left] at (-1.0,-3.0) {-\footnotesize{1.5}};
		\node [left] at (-1.0,-2.0) {-\footnotesize{1.0}};
		\node [right] at (1.80,-1.45) {-\footnotesize{0.72496}};
		\node [right] at (-0.55,1.15) {\footnotesize{no auto}};
		\node [right] at (1.00,-0.70) {\footnotesize{max}};
		\node [right] at (1.00,-0.95) {\footnotesize{SC}};
		\node [right] at (-1.45,-0.70) {\footnotesize{no}};
		\node [right] at (-1.45,-0.95) {\footnotesize{SC}};			
		
		\draw [dashed] (3,1.5) -- (6.8,1.5);
		\draw [dashed] (3.5,-2.75) -- (3.5,2.0);
		\draw [dashed] (6.5,-2.75) -- (6.5,2.0);
		\draw [dotted] (4.0,-2.75) -- (4.0,2.0);
		\draw [dotted] (3,-2.25) -- (6.8,-2.25);
		\fill[pattern=north west lines, pattern color=blue, opacity=0.7]  (4.0,-2.25) --(4.0,1.5) -- (6.5,1.5) -- (6.5,-2.25) -- (4.0,-2.25)  ; 
		\path [fill=red!45, fill opacity=0.6]   (6.5,1.5) -- (5.5,-3) -- (5.166,-3.1) -- (4.833,-2.9) -- (4.5,-3.0) -- (6.5,1.5); 
			
		\draw [<->, thick] (3,-0.1) -- (7,-0.1);
		
		\node [below] at (7.2,-0.10)  {$\mu_{1n}$};
		\node [right] at (4.65,1.65) {\footnotesize{no auto}};
		\node [right] at (6.50,-0.70) {\footnotesize{max}};
		\node [right] at (6.50,-0.95) {\footnotesize{SC}};
		
		\node [below] at (3.5,-2.80) {\footnotesize{0.2}};
		\node [below] at (4.0,-2.80) {\footnotesize{0.25}};
		\node [below] at (6.5,-2.80) {\footnotesize{0.5}};
		\node [left] at (7.3,1.5) {\footnotesize{0.5}};
		\node [left] at (7.45,-2.35) {-\footnotesize{0.72496}};
		
		\node [above] at (2.25,-2.75) {$\frac{x^*}{\overline{x}_c}=15.625$};
	\end{tikzpicture} 
	\caption[Region of Validity in $\mu_{1n}$-$\mu_{1m} $: Clean Cloud]{Same as Figure \ref{fig:LG_values10um} except the lower bound of the triangle in the figures extends to $(-\frac12,-15.125)$, and is representative of the situation when the threshold radius is 40$\mu$m and the mean cloud radius is 16$\mu$m.  \label{fig:LG_values16um} }
\end{center}	
\end{figure}
\end{center}

\begin{table}[htb!]
\begin{center}
\begin{tabular}{ c >{\centering\arraybackslash}m{1.75cm}  >{\centering\arraybackslash}m{5.5cm} >{\centering\arraybackslash}m{1.75cm} }
\hline
Parameter & Lower  & Condition & Upper  \\ 
\hline \hline
$\mu_{1n}$ & 0.34320 & --- & 0.49300 \\

\multirow{2}{*}{$ \mu_{1m} $ } &  -0.7250 & $0.34320 \le \mu_{1n} \le 0.42160$   &  7.8125$\mu_{1n}$-3.40625 \\

 &15.625$\mu_{1n}$-7.3125  & $0.42160 \le \mu_{1n} \le 0.49300$  &  7.8125$\mu_{1n}$-3.40625   \\
 
$\mu_{4np}$ & \multicolumn{3}{ c }{=0.25}    \\

\hline
\end{tabular} 
\caption[Stochastic Parameter Bounds: Clean Cloud]{The same as Table \ref{tab:ST_param10um} except with a mean cloud mass of 16$\mu$m.\label{tab:ST_param16um} }
\end{center}
\end{table}

\section{Numerical Simulations} \label{sec:NumSim}
The intersection of the rectangular (blue), `SB subset region,' and triangular (red), `region of validity' of parameter values form polygons.  Parameter values within these polygons both represent physically real conditions and are a function of Seifert and Beheng's parameters.  We examine the evolution of aggregate number concentration and mass by the stochastic bulk model by (i) varying the cloud-self collection and auto conversion parameters within the intersection of the `region of validity' and the `SB subset region,' (ii) observing the time required to convert 50\% of cloud droplet mass to rain droplets within this intersection, and (iii) comparing the ``50\% time" to results from the Linear Flux Method \cite{AB98}.  We show the effects of macroscopic cloud properties within the red-blue intersection of by calculating the initial rain droplet radius and the radius at the time of 50\% conversion.  Finally, sensitivity tests are performed w.r.t the time step and the rain self-collection parameter in Section \ref{sec:SensitivityTests}.

\subsection{Model Setup}
We perform simulations for two cloud types: polluted and clean.  The polluted clouds have an initial mean cloud droplet radius of 10 $\mu$m which is 239 droplets per cm$^{3}$ given the typical liquid water content of 1 g cm$^{-3}$.  Clean clouds have an initial mean cloud droplet radius of 16 $\mu$m and a corresponding droplet concentration of  58.3 cm$^{-3}$.  The third stochastic parameter $\mu_{\text{4np}}$, regulating rain self-collection is fixed at 0.25.  All simulations use 3$^{\text{rd}}$ order Adams-Bashforth with Improved Euler and AB2 for initialization.

The stochastic model is validated against results from a detailed evolution of the cloud droplet spectrum.  We use the Linear Flux Method by Bott (1998) with 70 bins that double every second bin.  The smallest droplet radius on the discretized spectrum is 1 $\mu$m and the largest is 3.251 mm.

\subsection{Results} \label{sec:EvolutionGraphs}
Three types of results are shown for each cloud type.  We present contour graphs of the 50\% conversion time that span relevant portions of the intersection of the the region of validity and the `SB subset region.'  The relevant portions are where changes in the 50\% conversion time as a function of the cloud self-collection and auto conversion parameters are noticeable.  This occurs where cloud self-collection is stronger and auto conversion is weaker, where $\mu_{1n}$ and $\mu_{1n}$ are near by slightly less than 0.50.  Within this same region in the ($\mu_{1n},\mu_{1n}$) space we present contour graphs showing how the mean rain radius, at the 50\% conversion time, changes as the stochastic parameters are varied.  Additionally we present another type of contour graph that shows how the mean rain radius at the first time step varies as a function of ($\mu_{1n},\mu_{1n}$).  The third result is a set of curves of mass-over-time graphs comparing the conversion of cloud to rain mass by the stochastic model, Seifert and Beheng's model, and the detailed Linear Flux Method \cite{AB98}.

\subsubsection{50\% Mass Conversion: Contour Graphs} \label{sec:Conversion}
\begin{figure}[t]
\begin{center}
	\begin{tikzpicture}[thick,scale=1.00, every node/.style={scale=0.75}] 
		\draw [<->, thick] (-1.8,-2.75) -- (2.0,-2.75);
		
		\draw [dashed] (-1.8,1) -- (1.85,1);
		\draw [dashed] (1,-2.80) -- (1,1.5);
		\draw [dotted] (0.75,-2.0) -- (1.85,-2.0);
		\draw [dotted] (0.75,0.75) -- (1.85,0.75);
		\draw [dotted] (0.75,-3.0) -- (0.75,-2.0);
		\draw [dotted] (-1.30,-3.0) -- (-1.30,-2.0);

		\path [fill=red!25] (1.0,1) -- (-1.30,-2) -- (-1.80,-2.00) -- (-2.00,-1.650) -- (-1.80,-1.35) -- (-2.0,-1.0) -- (-1.9,-0.8)  -- (1.0,1)  ;
		\fill[pattern=north west lines, pattern color=blue, opacity=0.7]  (-1.8,-2.0) -- (-2.0,-1.0) -- (-1.6,0.0) -- (-1.8,0.75) -- (0.75,0.75) -- (0.75,-2.0) -- (0.5,-2.0)  ; 		
		\node [right] at (2.0,-2.75)  {$\mu_{1n}$};
		
		\draw [solid, thin] (0.45,0.95) -- (0.90,0.95);
		\draw [solid, thin] (0.45,0.30) -- (0.45,0.95);
		\draw [solid, thin] (0.45,0.30) -- (0.90,0.30);
		\draw [solid, thin] (0.90,0.30) -- (0.90,0.95);

		\node [right] at (1.80,1.00) {\footnotesize{0.5}};
		\node [above] at (1.0,1.5) {\footnotesize{0.5}};
		\node [right] at (1.80,0.75) {\footnotesize{0.49829}};
		\node [right] at (1.80,-2.0) {\footnotesize{0.20094}};
		\node [below] at (-1.30,-3.0) {\footnotesize{0.49530}};
		\node [below] at (0.55,-3.0) {\footnotesize{0.49997}};
		\node [right] at (-0.55,1.15) {\footnotesize{no auto}};
		\node [right] at (1.00,-0.70) {\footnotesize{max}};
		\node [right] at (1.00,-0.95) {\footnotesize{SC}};
		
		\node[inner sep=0pt] (10xc40xs) at (6.15,-0.75)
    {\includegraphics[width=.60\textwidth,trim=40 200 35 185,clip]{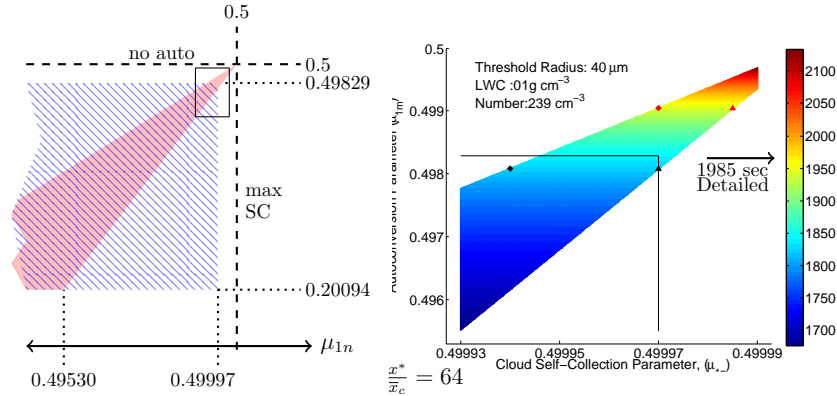}};
	
    		\draw [->, thick] (7.25,-0.25) -- (8.15,-0.25);
		\node [right] at (7.00,-0.60) {\footnotesize{Detailed}};
		\node [right] at (7.00,-0.40) {\footnotesize{1985 sec}};

		\node [above] at (3.50,-3.50) {$\frac{x^*}{\overline{x}_c}=64$};
	\end{tikzpicture} 
	\caption[Contour Plot on $\mu_{1n}$-$\mu_{1m} $ plane: 50\% Conversion time, ~~ Polluted Cloud]{The graph on the left is a blow-up of the two graphs in Figure \ref{fig:LG_values10um}.  The solid (red) region contains values of the cloud self-collection and auto conversion parameters that adhere to consistency of number and conservation of mass.  The patterned (blue) region contains values which reproduce Seifert and Beheng's bulk parameterization.  The thin box in the upper right is the region of the contour graph on the right, which shows that instant at which rain mass exceeds cloud mass as a function of the two stochastic parameters.  The solid lines in the contour plot correspond to the upper right edge of the patterned (blue) region in the plot on the left.  These graphs are representative of the situation when the threshold radius is 40$\mu$m and the mean cloud radius is 16$\mu$m.  Neither graphs in Figure \ref{fig:Contour_10um} are to scale.    \label{fig:Contour_10um} }
\end{center}	
\end{figure}

Contour graphs depicting the time needed to convert 50\% of cloud mass to rain mass are shown for simulations with an initial mean cloud radius ($\ol{x}_c$) of 10 $\mu$m (polluted) and 16$\mu$m (clean). The polluted simulation requires a cloud self-collection phase to grow the mean cloud droplet size large enough for  auto conversion to have a significant effect.  Therefore, the cloud self-collection parameter can be expected to be strong and the auto conversion parameter to be weaker as is shown by the location of the thin-lined black box in the upper corner of the left-hand graph in Figure \ref{fig:Contour_10um}.  

Both graphs in Figure \ref{fig:Contour_10um} are blow-ups of the graphs in Figure \ref{fig:LG_values10um}.    In the graph on the left, the solid (red) region contains values of the cloud self-collection and auto conversion parameters that adhere to consistency of number and conservation of mass.  The patterned (blue) region contains values which reproduce Seifert and Beheng's bulk parameterization.  This graph is not to scale, but identifies where the contour map is located relative to the graphs in Figure \ref{fig:LG_values10um}.  The two black line segments in the contour map correspond to the edge of the patterned (blue) area within the black box in the graph on the left.  Stochastic parameter values below and to the left of the black lines will reproduce Seifert and Beheng's bulk parameterization given the the bounds on their parameters $0 \le \nu \le 3$ and $0 \le \tau \le 1$.  Stochastic parameter values which are above and to the right of the black lines do not reproduce their bulk parameterization.

\begin{figure}[t]
\begin{center}
	\begin{tikzpicture}[thick,scale=1.00, every node/.style={scale=0.75}] 
		
		\draw [dashed] (-1.8,1) -- (1.85,1);
		\draw [dashed] (1,-2.80) -- (1,1.5);
		\draw [dotted] (0.65,0.95) -- (2.35,0.95);
		\draw [dotted] (0.65,-3.1) -- (0.65,-2.75);
		\draw [dotted] (-0.22,-3.10) -- (-0.22,-2.75);

		\path [fill=red!25] (1.0,1) -- (-0.30,-3) -- (-0.90,-2.90) -- (-1.50,-3.00) -- (-1.9,-2.9)  -- (1.0,1)  ;
		\fill[pattern=north west lines, pattern color=blue, opacity=0.7]  (-1.8,-2.95) -- (-2.0,-1.6) -- (-1.6,-0.3) -- (-1.8,0.95) -- (0.65,0.95) -- (0.65,-2.95)  -- (-0.15,-3.05) -- (-1.05,-2.85) -- (-1.8,-2.95)  ; 		
		\node [right] at (2.0,-2.75)  {$\mu_{1n}$};
		\draw [<->, thick] (-2.0,-2.75) -- (2.0,-2.75);
		
		\draw [solid, thin] (0.45,1.00) -- (0.80,1.00);
		\draw [solid, thin] (0.45,-0.60) -- (0.45,1.00);
		\draw [solid, thin] (0.45,-0.60) -- (0.80,-0.60);
		\draw [solid, thin] (0.80,-0.60) -- (0.80,1.00);

		\node [right] at (1.50,1.15) {\footnotesize{0.5}};
		\node [above] at (1.0,1.5) {\footnotesize{0.5}};
		\node [right] at (1.95,0.85) {\footnotesize{0.49650}};
		\node [below] at (-0.30,-3.1) {\footnotesize{0.46800}};
		\node [below] at (0.80,-3.1) {\footnotesize{0.49300}};
		\node [right] at (-0.55,1.15) {\footnotesize{no auto}};
		\node [right] at (1.00,-0.70) {\footnotesize{max}};
		\node [right] at (1.00,-0.95) {\footnotesize{SC}};
		
		\node[inner sep=0pt] (10xc40xs) at (6.15,-0.75)
    {\includegraphics[width=.60\textwidth,trim=40 200 35 185,clip]{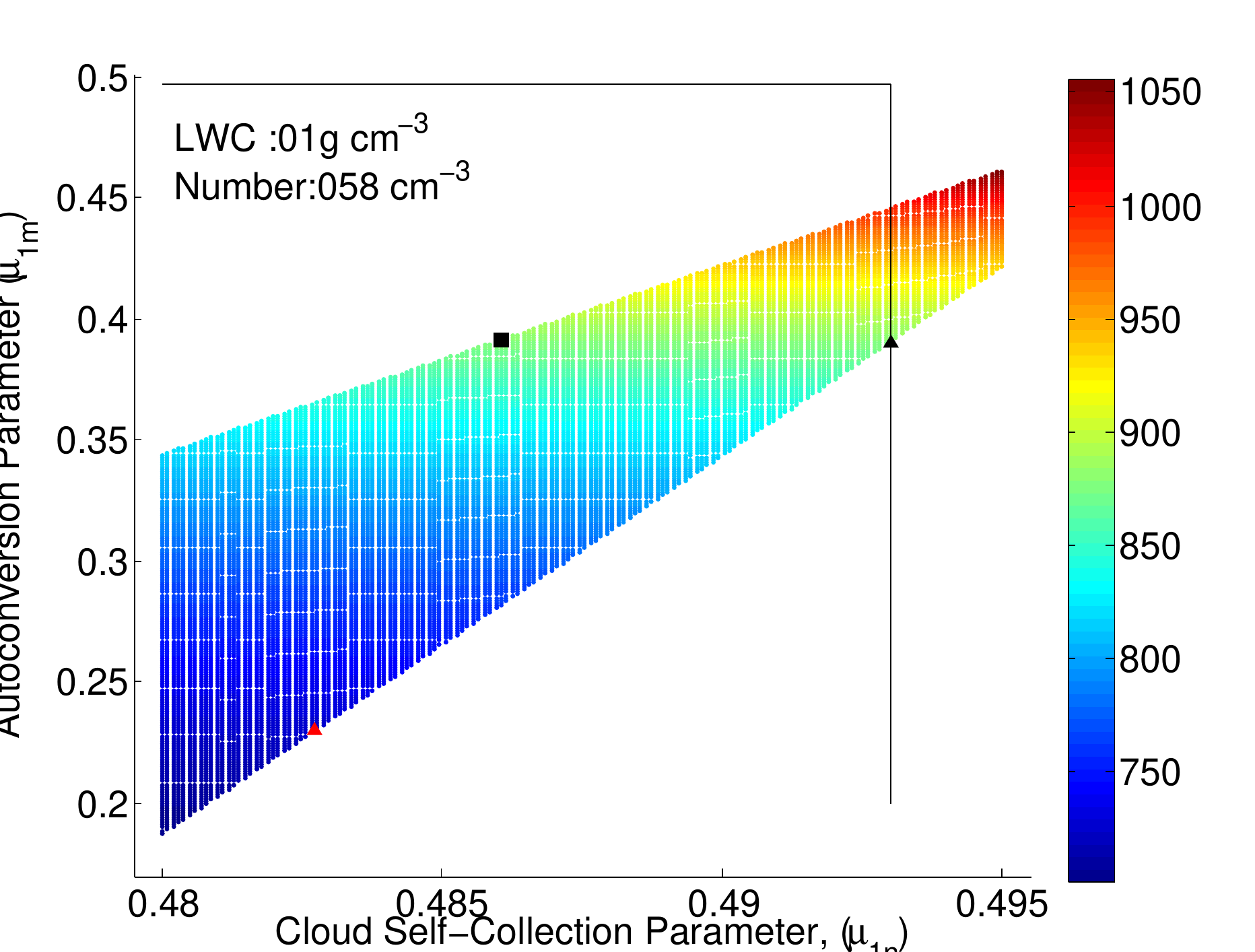}};
    
        		\draw [->, thick] (6.50,-2.40) -- (8.13,-2.40);
		\node [right] at (6.00,-2.05) {\footnotesize{Detailed}};
		\node [right] at (6.00,-2.25) {\footnotesize{721 sec}};
		
		\node [above] at (3.50,-3.50) {$\frac{x^*}{\overline{x}_c}=15.625$};
	\end{tikzpicture} 
	\caption[Contour Plot on $\mu_{1n}$-$\mu_{1m} $ plane: 50\% Conversion time, ~~ Clean Cloud]{These are the same as in Figure \ref{fig:Contour_10um} except the graph on the left is a blow-up of the two graphs in Figure \ref{fig:LG_values16um} and the mean cloud radius is 16$\mu$m.  The details related to the diamonds and square are explained in the text.  Neither graphs in Figure \ref{fig:Contour_16um} are to scale.   \label{fig:Contour_16um} }
\end{center}	
\end{figure}

\begin{figure}[thb!]
\centering{
	\includegraphics[width=.65\textwidth,trim=30 185 35 175,clip]{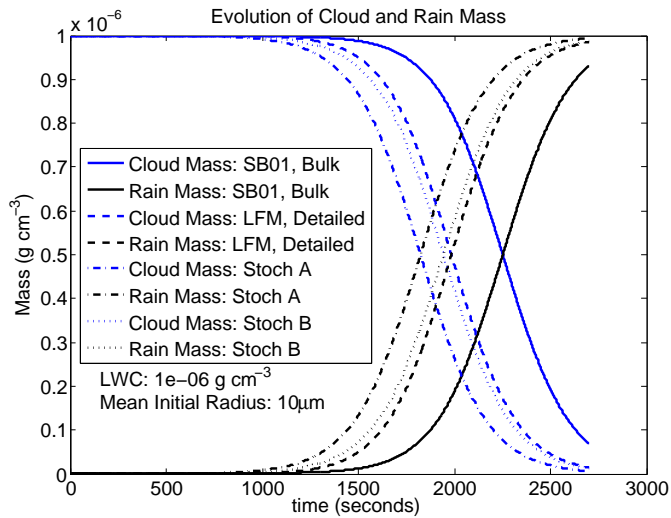} }
	\caption[Evolution Curves: Polluted Cloud]{The stochastic bulk parameterization is evolved using the parameters listed in the graph.  The parameter values are high lighted in Table \ref{tab:Evo_10um}.  Initial conditions: 1 g m$^{-3}$, 238.7 droplets cm$^{-3}$, and 10$\mu$m mean radius.  The evolution is compared to a detailed evolution of the kinetic collection equation using Bott's Linear Flux Method (1998) and Seifert and Beheng's bulk parameterization.  The stochastic parameterization evolves the aggregate mass faster than both Seifert and Beheng's and the detailed results, and is closer to the detailed results than Seifert and Beheng's parameterization.  The parameters for Stoch A are $\mu_{1n}$ =  0.49997, and $\mu_{1m}$ = 0.49808, and for Stoch B: $\mu_{1n}$ =  0.49997, and $\mu_{1m}$ = 0.49904.   \label{fig:Evo_SB01_ST15_10um} } 
\end{figure}

The contour map (on the right) in Figure \ref{fig:Contour_10um} shows how strongly the evolution is affected by small changes in the auto conversion parameter in this small corner in the region of permissible values.  The black triangle identifies the stochastic parameter values (0.49997, 0.49808) and the red diamond is  (0.49997, 0.49904).  These parameters convert 50\% of cloud droplet mass to rain droplet mass in 1824 and 1944 seconds, respectively.  The black diamond and the red triangle have the same \%50 mass conversion time as the corresponding marks of the same colour.  This is depicted in the contour graph by noting that contour lines are horizontal, i.e. mass conversion by the polynomial kernel is independent of the cloud self-collection parameter.  The evolution curves corresponding to these parameters are in Figure \ref{fig:Evo_SB01_ST15_10um}.  

The red marks are located outside of the region of stochastic parameter values that will reproduce Seifert and Beheng's bulk parameterization.  The stochastic parameters are constant throughout the evolution.  Seifert and Beheng's non-dimensional time parameter $\tau$ changes as the cloud matures.  The auto conversion parameter $\mu_{1m}$ can be considered as an average of possible auto conversion parameter values during the maturation of a cloud.  The $\tau$ function $\Phi_{\text{auto}}(\tau)$ is concave.  Jensen's inequality shows that the concave function of a mean gives an over estimate  \cite{VL01}, and thus unduly increasing the strength of the auto conversion parameter.  When the incidental strengthening of the auto conversion parameter is mitigated by relaxing the restrictions imposed by the bounds of $\nu$ and $\tau$ on the values of the stochastic parameters, the stochastic model produces results consistent with detailed results using stochastic parameters identified by the (red) diamond.

The contour graphs for simulations with an initial mean cloud radius of 16$\mu$m is shown in Figure \ref{fig:Contour_16um} with the thin black boxes and lines, and the solid (red) and patterned (blue) regions serving the same purpose as in Figure \ref{fig:Contour_10um}.  These graphs detail the regions presented in Figure \ref{fig:LG_values16um}.  The black triangle, and square, identify the stochastic parameter pairs (0.493013, 0.390833) and (0.486060, 0.391093), respectively.  These each convert 50\% of mixing ratio to rain mixing ratio in 880 seconds.  The evolution curves of both pairs are shown in Figure \ref{fig:Evo_SB01_ST15_16um}.  The red diamond identifies the stochastic parameter values (0.482682, 0.229417) and converts 50\% of mixing ratio to rain mixing ratio in 726 seconds..  A cloud with a  mean cloud radius of 16$\mu$m has more liquid water content contained larger cloud droplets.  A stronger (smaller valued) auto conversion parameter is expected, and is seen when comparing the results from the simulations with an initial droplet radius of 10$\mu$m.

\begin{figure}[thb!]
\centering{
	\includegraphics[width=.48\textwidth,trim=30 185 05 175,clip]{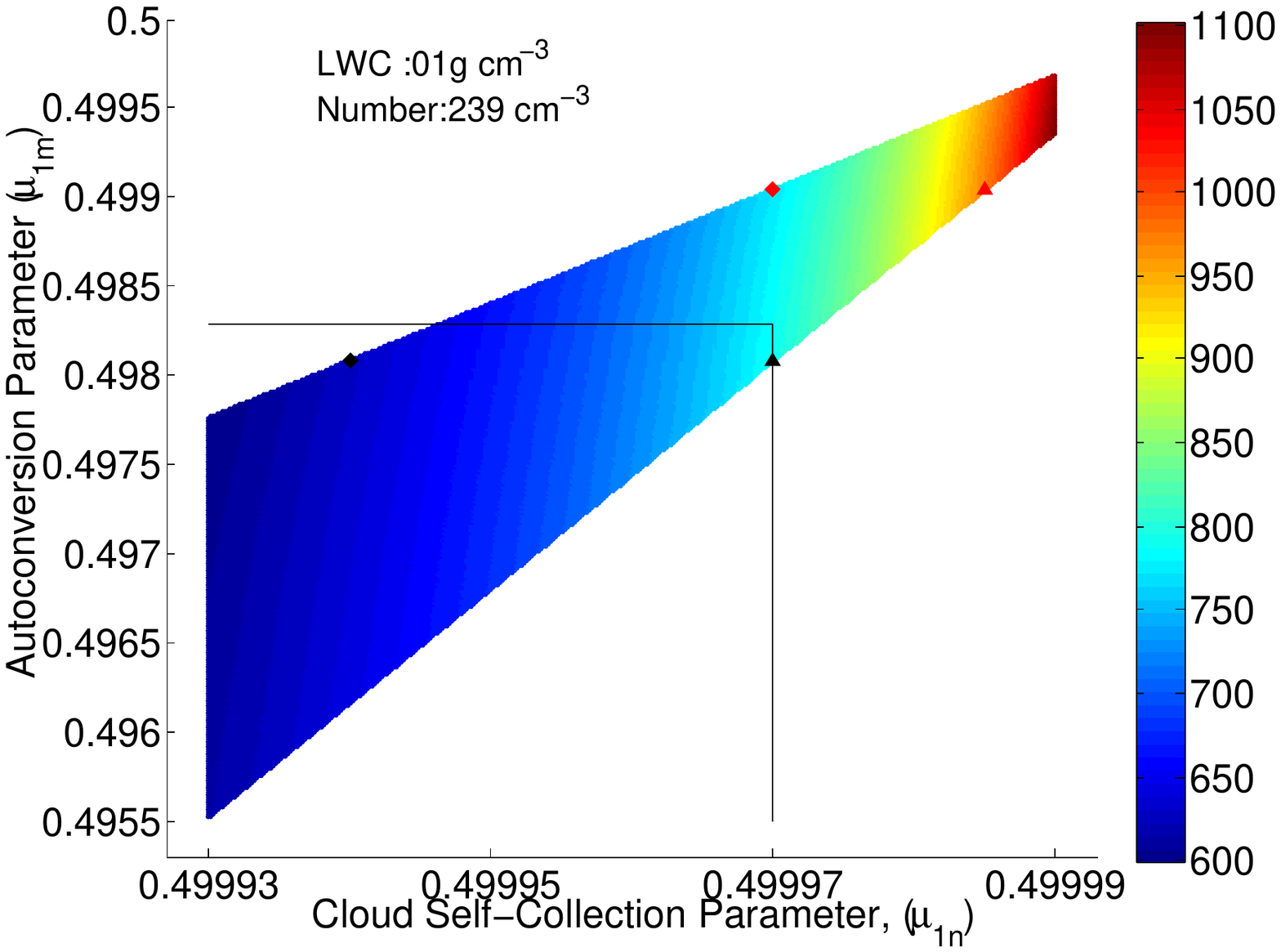} 
	\includegraphics[width=.48\textwidth,trim=30 185 05 175,clip]{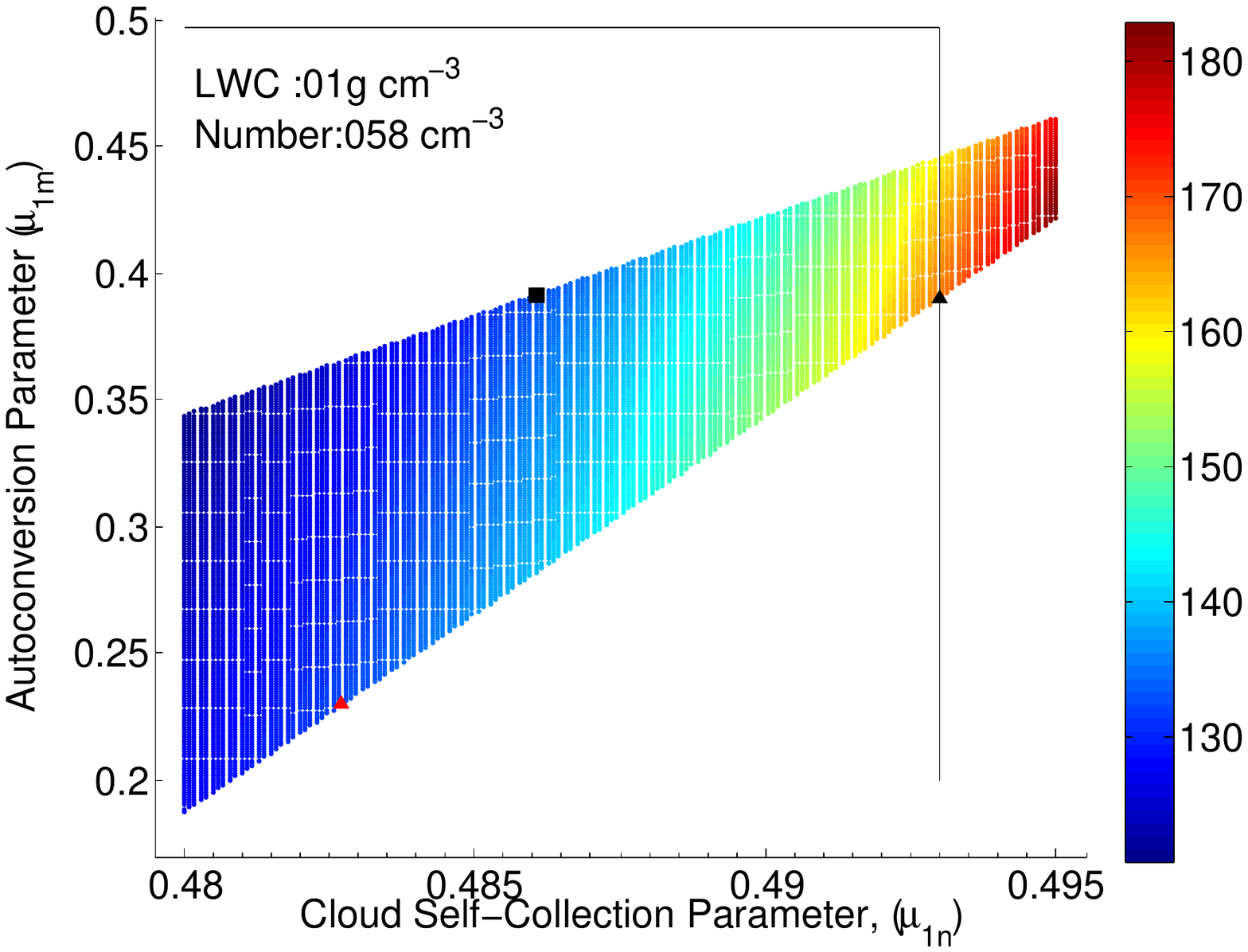} }
	\caption[Contour Plot on $\mu_{1n}$-$\mu_{1m} $ plane: Rain Radius, Polluted Cloud]{Mean rain radius at the time of 50\% mass conversion to rain droplets.  Initial conditions:  1 g m$^{-3}$ for both, and 239 (16 $\mu$m) and 58.3 droplets cm$^{-3}$ (16 $\mu$m) initial droplet concentration for the left and right graph, respectively.      \label{fig:RainRadius_SB01_ST15} } 
\end{figure} 

\subsubsection{Mean Rain Radius: Contour Graphs} \label{sec:MRRConGr}
The mean rain radius is an important complement to the 50\% conversion time metric in a zero-degree parcel cloud model.  The 50\% conversion time quantifies how quickly cloud water is redistributed from being contained in cloud droplets to forming rain droplets.  However, in a parcel model simulation the newly formed rain droplets are essentially ``still in the cloud," and have not started precipitating.  This is a significant difference between a parcel model and a 1-D rain shaft model.  By analyzing the conversion time in conjunction with mean rain droplet radius at 50\% conversion time, we can get an idea of the onset of precipitation for particular stochastic parameter values relative to other stochastic parameter values and relative to other bulk and detailed models.  A 1-D model would be appropriate when comparing to observational data.  However, the inclusion of observational data is beyond the scope of this work.

A larger mean rain droplet radius at the time of 50\% conversion will lead to a quicker onset of precipitation, and a smaller mean rain droplet radius at that time indicates a slower onset of precipitation.  The mean rain radius at the 50\% time is shown in Figure \ref{fig:RainRadius_SB01_ST15} for both the polluted (10 $\mu$m) case and the clean (16 $\mu$m) case.  The boundary of the `SB subset region' and the same triangular and square markers, identifying the location of particular parameter values, as shown in Figures \ref{fig:Contour_10um} and \ref{fig:Contour_16um} are replicated in the rain radius contour plot.  These two contour plots also show that there is a greater dependence, of the mean rain radius at the 50\% time of conversion, on the cloud self-collection parameter than on the auto conversion parameter.

\begin{figure}[tbh!]
\centering{
	\includegraphics[width=.48\textwidth,trim=30 185 05 175,clip]{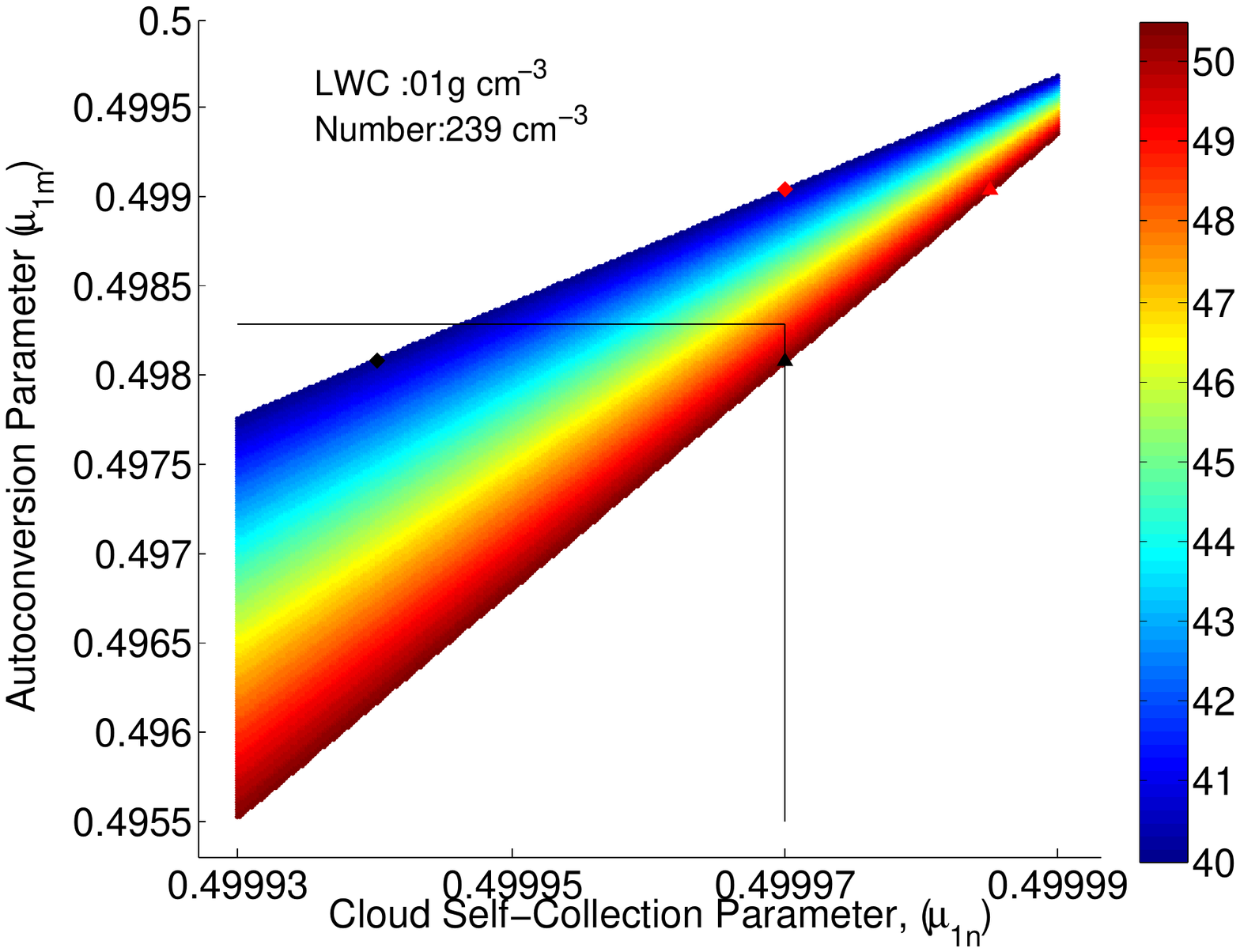} 
	\includegraphics[width=.48\textwidth,trim=30 185 05 175,clip]{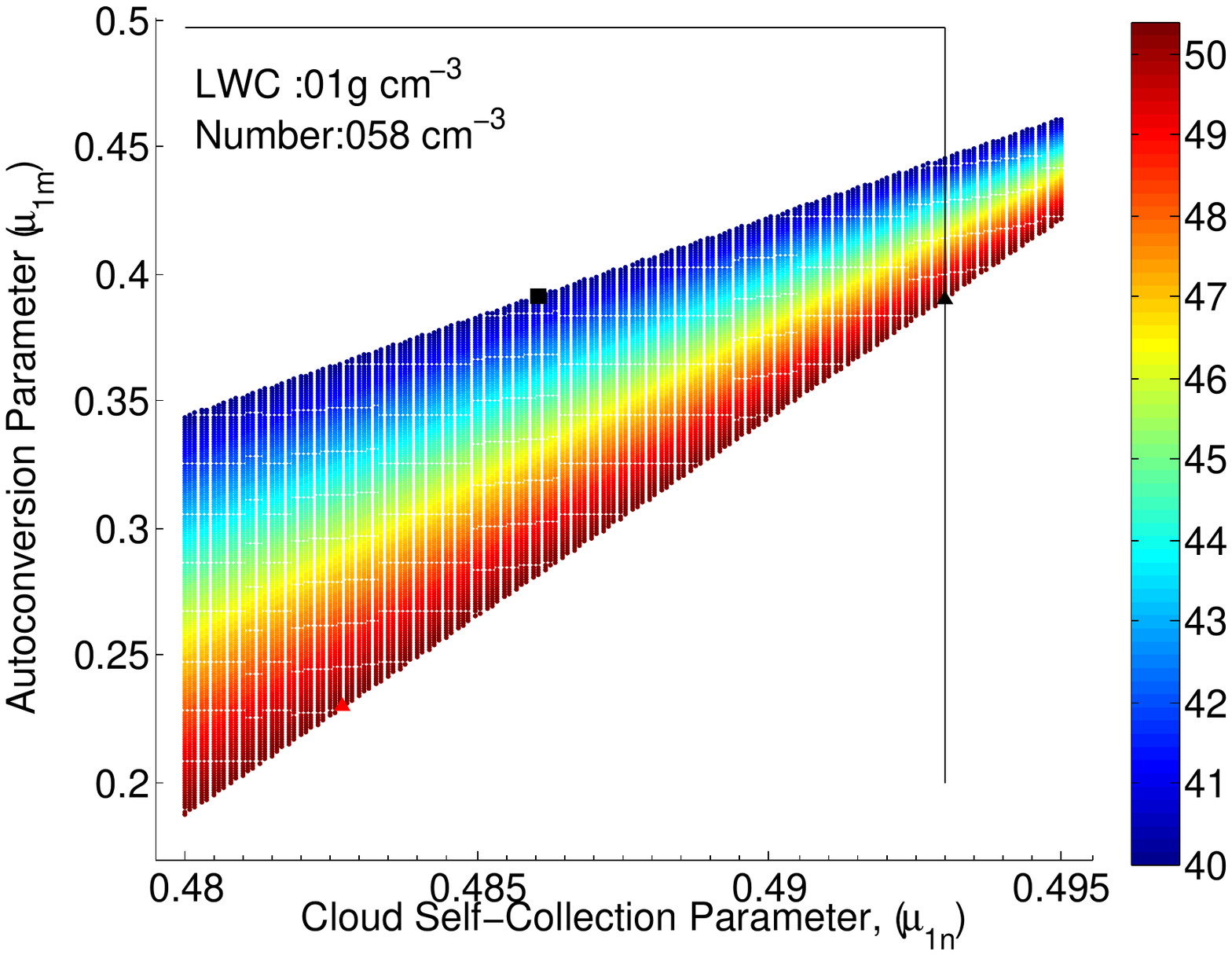} }
	\caption[Contour Plot on $\mu_{1n}$-$\mu_{1m} $ plane: Rain Radius, Clean Cloud]{Mean rain radius after the first time step.  Initial conditions:  1 g m$^{-3}$ for both, and 239 (16 $\mu$m) and 58.3 droplets cm$^{-3}$ (16 $\mu$m) initial droplet concentration for the left and right graph, respectively.   \label{fig:InitRadius_SB01_ST15} } 
\end{figure} 

The initial (at the first time step) mean rain radius is shown in the two contour plots in Figure \ref{fig:InitRadius_SB01_ST15}.  In both cases, the upper boundary has an initial mean rain radius of 40 $\mu$m which is equal to the threshold radius.  This boundary separates the pre-collision droplets that form a rain droplet during the auto conversion process from ones that form a cloud droplet during this particular collision process.  Since the latter contradicts the definition of auto conversion, this upper boundary separates physically allowable stochastic parameter values from ones that are not allowable.  At the lower boundary, the initial mean rain radius is 50.4 $\mu$m.  This boundary separates the pre-collision cloud droplets that form a rain droplet during the auto conversion process from ones that formed by pairs of rain droplets during this particular collision process.  Since the latter contradicts the definition of auto conversion, this lower boundary also separates physically allowable stochastic parameter values from ones that are not allowable.  The range of initial rain radii is important when using particular turbulent kernels in the stochastic bulk model.  

For instance, a tabulated turbulent kernel by \cite{MP08} has the greatest enhancement when smaller cloud droplets are coalescing with larger cloud droplets.  A turbulent kernel such as this would be best served by stochastic parameters along the upper boundary ($\ol{x_r}(\text{\footnotesize{$t$=1}})=40\mu$m) since the strongest enhancement of collisions occurs with the smallest of droplets, the predominate size of newly produced rain droplets be approximately equal to the size of the threshold droplet: $x^*=40\mu$m.

In contrast, a parameterized kernel by \cite{CF08} has stronger enhancement factors when droplets of nearly the same size coalesce and the greatest enhancements are when the largest of the cloud droplets coalesce with each other.  This type of turbulent kernel would be best represented by the stochastic bulk model when using parameters along the lower boundary  ($\ol{x_r}(\text{\footnotesize{$t$=1}})=50.4\mu$m).  Two cloud droplets with radii slightly less than that of the threshold droplet would produce a rain droplet with a radii of $50.4=(2*40^3)^{(1/3)}$.

\subsubsection{Evolution of Mass Conversion} \label{sec:Evolution}
Cloud and rain mixing ratios are evolved using three methods: (i) the new `SDE-based' stochastic bulk parameterization, (ii) the bulk parameterization by Seifert and Beheng (2001), and (iii) a detailed scheme using Bott's Linear Flux Method (LFM) from 1998.  The results from the detailed method are taken as the benchmark.  Mixing ratios in polluted clouds are evolved for 45 minutes, and clean clouds (larger initial mean droplets) for 30 minutes.  The stochastic parameter pairs chosen for these simulations are taken from the marker locations on the contour plots in Figures \ref{fig:Contour_10um} and \ref{fig:Contour_16um}.   

\begin{table}[tbh!]
\begin{center}
\begin{tabular}{ c >{\centering\arraybackslash}m{1.5cm}  >{\centering\arraybackslash}m{1.5cm} >{\centering\arraybackslash}m{2.5cm}  }
\hline
\multirow{2}{*}{($\mu_{1n}$, $\mu_{1m}$)} &  \multicolumn{2}{ c }{$q_c = q_r$}  & $t=1$ second   \\ 
 & seconds & $\ol{x}_r$ $(\mu$m) &  $\ol{x}_r$ $(\mu$m)   \\
\hline \hline
(0.49994008, 0.49808267) & 1824 & 628.2 & 40.0 \\

(0.49997003, 0.49808213) & 1824 & 791.5 & 50.4  \\

(0.49997003, 0.49904107) & 1944 & 773.9 & 40.0 \\

(0.49998501, 0.49904053) & 1944 & 975.2 & 50.4  \\

Bulk: Seifert \& Beheng (2001) & 2253 & 151.3 & 39.6  \\

Detailed: Linear Flux Method & 1985 &  &   \\

\hline
\end{tabular} 
\caption[Evolution Results: Polluted Cloud]{The first column gives the parameter values used to acquire the data in the remaining columns.  The second and third columns give the time (seconds) and the mean rain radius ($\mu$m) at the first time step that rain mixing ratio exceeds cloud mixing ratio.  The fourth column gives  the mean rain radius at the first time step.  The last column gives the time (seconds) at which the mean rain radius first equals the threshold radius.  The four parameter pairs correspond to the four markers in Figure \ref{fig:Contour_10um}   \label{tab:Evo_10um} }
\end{center}
\end{table}

For the polluted cloud case, the stochastic parameter pairs used to generate the two stochastic curves in Figure \ref{fig:Evo_SB01_ST15_10um} are in the second and third rows of Table \ref{tab:Evo_10um}.  The first and fourth rows show simulations that generate the same 50\% conversion time, but have different mean rain radii at the 50\% time.  The stochastic parameters that produce a 50\% conversion time of 1944 seconds were specifically chosen to match the detailed results of 1985 seconds.  Parameters that exactly reproduced the time of 1985 seconds we're available.  However, the evolution curves were indistinguishable.  So, we chose values that highlighted how closely the curves matched.  Seifert and Beheng's conversion rate was considerably slower.  However, the retarded growth by Seifert and Beheng's parameterization is consistent with Franklin's (2008) analysis that their model of auto conversion underestimates the auto conversion rate \cite{CF08}.  

\begin{figure}[thb!]
\centering{
	\includegraphics[width=.65\textwidth,trim=30 185 05 175,clip]{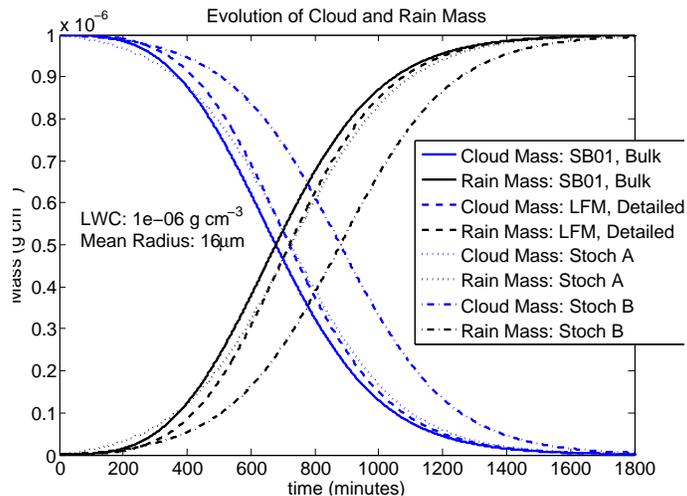} }
	\caption[Evolution Curves: Clean Cloud]{Same as Figure \ref{fig:Evo_SB01_ST15_10um} except for the initial conditions:  1 g m$^{-3}$, 58.3 droplets cm$^{-3}$, 16 $\mu$m mean radius.  The cloud self-collection and auto conversion parameters for Stoch A are $\mu_{1n}$ =  0.482682, and $\mu_{1m}$ = 0.229417 and for Stoch B are $\mu_{1n}$ =  0.493013, and $\mu_{1m}$ = 0.390833.  The latter pair generate an evolution curve that adhere to all physical constraints and lies within the region generated by the bound of Seifert and Beheng's parameters  $\nu$ and $\tau$.   \label{fig:Evo_SB01_ST15_16um} } 
\end{figure}

Referring to Figure \ref{fig:Contour_10um}, we see that the intersection of the `SB-subset region' and the `region of validity' is a truncation of the triangle due to the limits imposed by the values of $\nu$ and $\tau$ that Seifert and Beheng chose.  When recreating their model with the stochastic parameterization, the space of stochastic parameter values was created using the maximums and minimums of $\nu$ and $\tau$.  However, $\tau$ changed during the evolution while the stochastic parameters remained constant.  This arbitrary limitation calls for the relaxation of the truncation on the triangle due to the bounds on $\mu_{1n}$ and $\mu_{1m}$ imposed by $\nu$ and $\tau$ because the $\tau$ function is concave and by Jensen's inequality overestimates the auto conversation terms \cite{VL01}.  To compensate for this acceleration, values of $\mu_{1n}$ and $\mu_{1m}$ which inhibit the evolution are allowed.  These values are stochastic parameter pairs of ($\mu_{1n},\mu_{1m})$, such that $\mu_{1n} > 0.49997$ and $\mu_{1m} > 0.49829$ are allowed.  

Choosing stochastic parameters that adhere to the threshold droplet restriction, i.e. contained within the (solid, red) triangle in Figure \ref{fig:LG_values10um}, but not necessarily adhering to the derived restrictions on the auto conversion parameter, produces a bulk evolution curve more reflective of the detailed benchmark as shown in Figure \ref{fig:Evo_SB01_ST15_10um}.   Figure \ref{fig:Contour_10um} plots the locations of the coordinates ($\mu_{1n}$ $\mu_{1m}$) used to generate the results of the mean stochastic parameterization in Figure \ref{fig:Evo_SB01_ST15_10um} relative to the linear boundaries given by the relationship of auto conversion and the droplet threshold.


\begin{table}[h!]
\begin{center}
\begin{tabular}{ c >{\centering\arraybackslash}m{1.5cm}  >{\centering\arraybackslash}m{1.5cm} >{\centering\arraybackslash}m{2.5cm}  }
\hline
\multirow{2}{*}{($\mu_{1n}$, $\mu_{1m}$)} &  \multicolumn{2}{ c }{$q_c = q_r$}  & $t=1$ second   \\ 
 & seconds & $\ol{x}_r$ $(\mu$m) &  $\ol{x}_r$ $(\mu$m)    \\
\hline \hline
(0.482682, 0.229417) & 726 & 132.4  & 50.4  \\

(0.493013, 0.390833) & 880 & 167.4 & 50.4  \\

(0.486060, 0.391093) & 880 & 132.9  & 40.0  \\

Bulk: Seifert \& Beheng (2001) & 678 & 72.3 & 39.6  \\

Detailed: Linear Flux Method & 721 &  &   \\

\hline
\end{tabular} 
\caption[Evolution Results: Clean Cloud]{The first column gives the parameter values used to acquire the data in the remaining columns.  The second and third columns give the time (seconds) and the mean rain radius ($\mu$m) at the first time step that rain mixing ratio exceeds cloud mixing ratio.  The fourth column gives  the mean rain radius at the first time step.  The last column gives the time (seconds) at which the mean rain radius first equals the threshold radius.  The three parameter pairs correspond to the three markers in Figure \ref{fig:Contour_16um}   \label{tab:Evo_16um} }
\end{center}
\end{table}

For the clean cloud case, the stochastic parameter pairs used to generate the two stochastic curves in Figure \ref{fig:Evo_SB01_ST15_16um} are in the first two rows of Table \ref{tab:Evo_16um}. The evolution curve, `Stoch A,' is detailed in the first row of Table \ref{tab:Evo_16um} and closely matches the 50\% conversion time of the detailed simulation.  For the initial 5-7 minutes the conversion of cloud water to rain water is excessive.  At approximately 400 seconds the rate of conversion, as indicated by the slope of the respective graphs, is less than the rate for the detailed simulation.  The slower conversion rate of the stochastic parameterization persists for the duration of the evolution.

When $\ol{x}_c=16\mu$m cloud mass is closer to the cloud-rain threshold.  To accurately model this initial condition a more aggressive auto conversion parameter is needed.  Consequently, values of the auto conversion parameter $\mu_{1m}$ need to be smaller and are situated more towards the centre of the quadrilateral region.  The preferred stochastic value relative the region on which one can take values is shown in Figure \ref{fig:Contour_16um}.  This is in contrast to a more polluted initial condition such as when $\ol{x}_c=10\mu$m.  In this polluted situation the cloud mass is further from the cloud-rain threshold and is more accurately modelled by a weaker auto conversion parameter, and weaker to the extant that the concavity bias of Seifert and Beheng's non-dimensional time parameter places an excessive restriction on the cloud self-collection and auto conversion stochastic parameters as is shown by the diamond (red) marker in Figure \ref{fig:Contour_10um} being outside the patterned (blue) region.

\subsection{Sensitivity Tests}  \label{sec:SensitivityTests}
Two types of sensitivity tests were performed.  The effect that the time step had on the time required to convert 50\% of cloud droplets to rain droplets and the rain droplet radii at that time.  The third sensitivity test examines the effect that the rain number parameter has on these metrics.  Both tests were performed using an initial mean cloud radius of 10 $\mu$m, and cloud self-collection and auto conversion parameters of $(\mu_{1n},\mu_{1m})$=(0.49997,0.49808).

As expected, the rain number stochastic parameter $\mu_{4np}$ only effects the evolution of rain number.  When $\mu_{4np}$ is varied from 0.10 to 0.40, the mean rain radius varies by 12.7\%.

\begin{table}[htb!]
\begin{center}
\begin{tabular}{ c >{\centering\arraybackslash}m{1.5cm}  >{\centering\arraybackslash}m{1.5cm}  }
\hline
\multirow{2}{*}{$\mu_{4np}$} &  \multicolumn{2}{ c }{$q_c = q_r$}    \\ 
 & seconds & $\ol{x}_r$ ($\mu$m)    \\
\hline \hline
0.40 & 1824 & 742.7  \\

0.30 & 1824 & 775.0  \\

0.25 & 1824 & 791.6    \\
 
0.20 &  1824 &  808.5  \\
 
0.10 &  1824 &  843.3  \\
 
\hline
\end{tabular} 
\caption[Sensitivity: Rain Self-Collection Parameter]{Varying the rain self-collection parameter show little effect on rain droplet size.  When $\mu_{4np}$ varies from 0.10 to 0.40, the mean rain radius at the time that half of the liquid water content is converted to rain varies by 100.6$\mu$m (12.7\%). \label{tab:SensitivityRNC} }
\end{center}
\end{table}

\begin{table}[htb!]
\begin{center}
\begin{tabular}{ c >{\centering\arraybackslash}m{1.5cm}  >{\centering\arraybackslash}m{1.5cm}  }
\hline
timestep &  \multicolumn{2}{ c }{$q_c = q_r$}    \\ 
seconds & seconds & $\ol{x}_r$ ($\mu$m)    \\
\hline \hline
120 & 1920 & 797.6  \\

60 & 1860 & 818.0  \\

30 & 1830 & 795.9  \\



10 & 1830 & 800.1  \\


1 & 1824 & 791.6    \\
 
 
0.1  &  1823.6 &  791.0  \\
 
\hline
\end{tabular} 
\caption[Sensitivity: Time Step]{Varying the time step has a negligible effect on the time when rain mass exceeds cloud or the mean rain mass at that instant.  The mean rain radius at the 50\% time varies by only 4.9$\mu$m (0.62\%) between 0.1 and 30 seconds.   \label{tab:SensitivityTimestep} }
\end{center}
\end{table}

Varying the time step from 0.1 seconds to 30 seconds has a negligible effect on the time required to convert half of the initial cloud mass to rain droplets (column two in Table \ref{tab:SensitivityTimestep}).  The mean rain radius at the time that half of the liquid water content is converted to rain varies by 4.9$\mu$m (0.62\%) between time steps of 0.1 and 30 seconds.  No advantage is gained by reducing the time step smaller than the 1 second used in the simulations in this paper.  It may be possible to relax the time step to 30 seconds without significantly compromising accuracy.  It should be noted that at 50 minutes of evolution the mean rain radius was greater than half a centimetre indicating that the omission of rain removal and droplet breakup were effecting the results.

\section{Conclusion} \label{sec:Conclu}
Assumptions and simplifications are necessary to produce computationally affordable parameterizations that represent cloud microphysical processes.  Parameterizations of cloud microphysical processes over the past forty-five years have made assumptions regarding the droplet size distribution \cite{AS01,YL04,YL06}.  Many parameterizations depend on ad-hoc parameters \cite{MK00,AS01,YL04,YL06,CF08}.  The stochastic bulk parameterization of collision and coalescence studied here does not rely on any particular distribution of the droplet size spectrum, but rather assumes that a distribution exists and it has a mean.  All of the parameters in the stochastic parameterization have physical meaning, and their values can be recovered from data.  The parameters represent the first moments of time series of aggregates of fluctuations and of product fluctuations.  These are fluctuations from the state space mean of mixing ratio and number over defined portions of the droplet size spectrum.  Recovering values of that data can be done graphically because the stochastic bulk parameterization sufficiently constraints the auto conversion parameters.  Values can also be recovered from detailed simulations, which can be deterministic or stochastic, such as Bott's computationally efficient bin-based method \cite{AB98} or the more computationally demanding Gillespie-See$\beta$elberg algorithm \cite{DG75b,MS96} as will be shown in a subsequent paper.  

The stochastic bulk parameterization of collision and coalescence is independent of any assumed droplet size distribution and yet retains the flexibility of applying any collision kernel to the resultant set of coupled ordinary differential equations.  The flexibility of using any collision kernel in the stochastic parameterization comes at the expense of retaining only the zeroth order term in the 2D Taylor expansion of the collision kernel centred at either the cloud, or rain, mean mass.  This flexibility requires that the value of the collision kernel be computed, or retrieved from a look-up table, while the stochastic parameterization is being utilized by a climate model.  

Stochastic parameters can be chosen that reflect particular cloud characteristics like type (maritime vs. continental and clean vs. polluted) and age.  This is particularly important because now a single bulk model of collision and coalescence can be used without the restriction of needing the same droplet size distribution for every cell in a climate model.  The new stochastic bulk model gives the flexibility of choosing different stochastic parameters for different cloud types.  Also, different cells in a climate model can use stochastic parameters associated with various collision kernels depending on the intensity and type of turbulence found in the clouds in that climate model cell.

We used the relationship of auto conversion to threshold droplet mass to define `regions of validity' for the auto conversion ($\mu_{1m}$) and the cloud self-selection ($\mu_{1n}$) parameters such that the selection of values for these parameters would cause the auto conversion terms in the stochastic bulk model to produce a rain droplet when two cloud droplets coalesced.  This restriction of the auto conversion parameter was only possible because each of the four occurrences of an auto conversion term  in the stochastic bulk model was derived independently of every other auto conversion term.  To the best of the authors' knowledge, no other bulk cloud microphysics model of collision and coalescence was able to accomplish this physics-based restriction on the auto conversion term because every other bulk model only derived one auto conversion term and then made the other three auto conversion terms to be functions of the derived term.

Other parameterizations have been restricted to a specific kernel \cite{AS01,YL04,YL06,CF08}.   The kernel in Franklin's parameterization was dependent on the turbulent strength, but limited in the range of turbulent kinetic energy used in that kernel.  The stochastic bulk parameterization can be used with a variety of kernels without any further derivations.  The value of the kernel at the mean cloud radius and mean rain radius is used in the stochastic bulk parameterizations.  Seifert and Beheng's \cite{AS01} parameterization used a kernel that contained both mean radii, and could be directly applied to the resultant stochastic derivation.  This application changed the powers of the evolving quantities in the stochastic equations, and the resultant powers of the evolved quantities matched the ones in Seifert and Beheng's differential equations. 

Moreover, physical constraints based on the physics of collision and coalescence processes were used to define a ``region of validity" for the stochastic parameters.  We matched both models term for term and solved for the stochastic parameters $\mu_{1m}$ and $\mu_{1n}$ as functions of their ad-hoc parameter.  We defined an `SB subset region' in the 2-D ($\mu_{1m},\mu_{1n}$) space by applying the bounds that they gave for their parameters.  Their bulk model is restricted to a piece-wise polynomial collision kernel, while ours retains the flexibility to use a variety of collision kernels.  Both models had precisely the same dependence on cloud and rain mixing ratio and number concentration.

For given values of the three stochastic parameters, the new stochastic parameterization produced results that closely matched results from detailed simulations that used Bott's LFM when the same initial conditions were used.  This was done for each of polluted and clean cloud types.  A simple metric of the time required to convert half of the cloud mixing ratio to rain mixing ratio was used to validate the results of the stochastic bulk model.   It is demonstrated here that while adequate parameter values can be found for the stochastic model, they can depend heavily of the initial mean cloud radius, i.e, the environmental conditions or the cloud's age. Moreover, for the two experiments considered here, despite the utilization of an ad hoc adjustment function to correct of excessive accretion and auto-conversion rates, Seifert and Beheng's model either overestimates or underestimates the time at which 50\% of the cloud mass is converted into rain.  Thus, the stochastic parameters that are found to faithfully reproduce the detailed KCE results can sometimes lies outside the `SB subset region.'  This is essence demonstrates the usefulness of the stochastic bulk parameterization.   As alternative to their ad-hoc auto conversion and accretion adjustment functions, Markov jump processes can be used to sample the ``region of validity," with the jump processes conditional on environmental conditions  (e.g. clean v.s. polluted, marine v.s. continental) and the age of the cloud \cite{BK10}. 

The sensitivity of the "50\% conversion time" was examined as a function of the 2D $\mu_{1n}$-$\mu_{1m}$ space.  This conversion time metric was sensitive to small changes in the auto conversion $\mu_{1m}$ parameter near the values that would reproduce the evolution curve produced by the detailed method, but showed very little sensitivity elsewhere in the domain of physically permissible values.  The conversion time metric showed little sensitivity to the cloud self-collection $\mu_{1n}$ parameter anywhere on this domain for a given value of the rain self-collection $\mu_{4np}$ parameter.  The mean rain radius at the time of 50\% conversion was sensitive to the value of the rain self-collection parameter.  Both the conversion time and the mean rain radius at 50\% conversion time showed very little sensitivity to changes in the time step in the range 0.1-30 seconds.  Subsequent research in the development of the stochastic bulk parameterization will apply hydrodynamic and turbulent kernels for the purpose of developing a methodology that models cloud microscopic processes using only cloud macroscopic information.


The intersection of the ``region of validity" and the ``SB subset region" defines a set of values for the stochastic parameters of cloud self-collection and auto conversion that both adheres to physical constraints and reproduces the bulk model by Seifert and Beheng (2001).  We validated the stochastic bulk model by finding stochastic parameter values that very closely reproduced the 50\% mass conversion time and the evolution curves.  This was done for the cases of a polluted cloud (239 cm$^{-3}$) and a clean cloud (58.3 cm$^{-3}$).  The model was further validated for both cloud types by noting that the mean rain radius at the first time step had physically meaningful bounds related to the process of auto conversion, i.e. the bounds on the rain radius at the first time step was dependent on the fact that the process of auto conversion produces one rain droplet from two cloud droplets.  

\textbf{Acknowledgements}
This research is part of D. Collins's Ph.D. thesis.  The research of B. Khouider is partly supported by a grant from the Natural Sciences and Engineering Research Council of Canada.  D. Collins's fellowship is partly funded through this grant.

\bibliographystyle{spmpsci}      
\bibliography{mybib_P2}   

\end{document}